\documentclass[%
 aip,
 amsmath,amssymb,
 reprint,%
]{revtex4-1}
\usepackage{graphicx}% Include figure files
\usepackage{subfigure}
\usepackage{dcolumn}% Align table columns on decimal point
\usepackage{bm}% bold math
\usepackage{color}
\usepackage[T1]{fontenc}
\usepackage{mathptmx}
\usepackage{multirow}
\usepackage{array}
\usepackage{float}
\usepackage{etoolbox}
\usepackage{booktabs}
\usepackage{url}
\usepackage{CJKutf8}
\usepackage{pifont}
\usepackage{epstopdf, epsfig}
\usepackage{graphicx}
\usepackage{longtable}
\usepackage[
    colorlinks=true,
    linkcolor=blue,
    urlcolor=blue,
    citecolor=blue
]{hyperref}

\usepackage{titlesec}
\titleclass{\subsubsubsection}{straight}[\subsection]
\newcounter{subsubsubsection}[subsubsection]
\renewcommand\thesubsubsubsection{\thesubsubsection.\alph{subsubsubsection}}
\titleformat{\subsubsubsection}
 {\normalfont\normalsize\bfseries}{\thesubsubsubsection}{1em}{}
\titlespacing*{\subsubsubsection}
{0pt}{3.25ex plus 1ex minus .2ex}{1.5ex plus .2ex}

\begin{document}
%\begin{CJK*}{UTF8}{gbsn}
\preprint{AIP/123-QED}

%\title{Multiscale modeling and simulation of thermodynamic nonequilibrium effects in three-dimensional high-speed compressible flows: Based on the discrete Boltzmann method}

\title{Effects of reflection distance on Richtmyer--Meshkov instability in the reshock process: A discrete Boltzmann study}

\author{Huilin Lai \begin{CJK*}{UTF8}{gbsn} (赖惠林)\end{CJK*}}
%\thanks{Corresponding author: hllai@fjnu.edu.cn}
\affiliation{School of Mathematics and Statistics, Key Laboratory of Analytical Mathematics and Applications (Ministry of Education), Fujian Key Laboratory of Analytical Mathematics and Applications (FJKLAMA), Center for Applied Mathematics of Fujian Province (FJNU), Fujian Normal University, 350117 Fuzhou, China}

\author{Chuandong Lin \begin{CJK*}{UTF8}{gbsn} (林传栋) \end{CJK*}}
\thanks{Corresponding author: linchd3@mail.sysu.edu.cn}
\affiliation{Sino-French Institute of Nuclear Engineering and Technology, Sun Yat-sen University, Zhuhai 519082, China}
\affiliation{Key Laboratory for Thermal Science and Power Engineering of Ministry of Education, Department of Energy and Power Engineering, Tsinghua University, Beijing 100084, China}

\author{Demei Li \begin{CJK*}{UTF8}{gbsn} (李德梅) \end{CJK*}}
\thanks{Corresponding author: dmli079@fjnu.edu.cn}
\affiliation{School of Mathematics and Statistics, Key Laboratory of Analytical Mathematics and Applications (Ministry of Education), Fujian Key Laboratory of Analytical Mathematics and Applications (FJKLAMA), Center for Applied Mathematics of Fujian Province (FJNU), Fujian Normal University, 350117 Fuzhou, China}

\author{Tao Yang \begin{CJK*}{UTF8}{gbsn} (杨涛) \end{CJK*}}
\affiliation{School of Mathematics and Statistics, Key Laboratory of Analytical Mathematics and Applications (Ministry of Education), Fujian Key Laboratory of Analytical Mathematics and Applications (FJKLAMA), Center for Applied Mathematics of Fujian Province (FJNU), Fujian Normal University, 350117 Fuzhou, China}

\author{Yanbiao Gan \begin{CJK*}{UTF8}{gbsn} (甘延标) \end{CJK*}}
\affiliation{Hebei Key Laboratory of Trans-Media Aerial Underwater Vehicle, School of Liberal Arts and Sciences, North China Institute of Aerospace Engineering, Langfang 065000, China.}

\author{Lingyan Lian \begin{CJK*}{UTF8}{gbsn} (连玲艳) \end{CJK*}}
\affiliation{School of Mathematics and Statistics, Key Laboratory of Analytical Mathematics and Applications (Ministry of Education), Fujian Key Laboratory of Analytical Mathematics and Applications (FJKLAMA), Center for Applied Mathematics of Fujian Province (FJNU), Fujian Normal University, 350117 Fuzhou, China}

\author{Aiguo Xu \begin{CJK*}{UTF8}{gbsn} (许爱国) \end{CJK*}}
\affiliation{National Key Laboratory of Computational Physics, Institute of Applied Physics and Computational Mathematics, P. O. Box 8009-26, Beijing 100088, P.R.China}
\affiliation{National Key Laboratory of Shock Wave and Detonation Physics, Mianyang 621999, China}
\affiliation{HEDPS, Center for Applied Physics and Technology, and College of Engineering, Peking University, Beijing 100871, China}
\affiliation{State Key Laboratory of Explosion Science and Safety Protection, Beijing Institute of Technology, Beijing 100081, China}

\date{\today}% It is always \today, today,

\begin{abstract}
The Richtmyer--Meshkov (RM) instability occurs when a perturbed interface between two fluids undergoes impulsive acceleration due to a shock wave. In this paper, a numerical investigation of the RM instability during the reshock process is conducted using the two-component discrete Boltzmann method. The influence of reflection distance on the RM instability, including both hydrodynamic and thermodynamic non-equilibrium effects, is explored in detail. The interaction time between the reflected shock wave and the material interface varies with different reflection distances. Larger reflection distances lead to a longer evolution time of the material interface before reshock, resulting in more complex effects on the interface deformation, the mixing extent of the fluid system, and non-equilibrium behaviors after reshock. Additionally, while the reflection distance has a minimal impact on mixing entropy before the secondary impact, a significant difference emerges after the secondary impact. This suggests that the secondary impact enhances the evolution of the RM instability. Furthermore, non-equilibrium behaviors or quantities exhibit complex dynamics due to the influence of the transmitted shock wave, transverse waves, rarefaction waves, material interfaces, and dissipation/diffusion processes.
\end{abstract}

\maketitle
%\end{CJK*}

\section{\label{sec:level1} Introduction}

The Richtmyer--Meshkov (RM) instability primarily arises from the baroclinic effect, which is caused by the misalignment between the pressure gradient and the density gradient across the material interface \cite{brouillette2002richtmyer}. This baroclinic effect results in the development of perturbation, which in turn enters the linear phase, nonlinear stage, and finally enters the turbulent stage. In addition, the RM instability plays an important role in processes such mixing and combustion in supersonic ramping engine and laser-driven inertial confinement fusion (ICF) \cite{zhou2025instabilities}, and generally exists in nature and engineering applications, such as aerospace \cite{zhou2024hydrodynamic}, weapons implosion \cite{schill2024suppression} and supernova explosion \cite{abarzhi2024perspective}, etc. In ICF, there exists mixing between the capsule material and fuel due to the RM instability, which prohibits the acquisition of useful production for power generation from fusion reactions \cite{li2024microphysics}. On the contrary, the RM instability is helpful in combustion systems, as it can enhance the mixing of the fuel and the oxidizer in the supersonic ramping engine \cite{chou2024computational}. Therefore, an insightful study of the RM instability holds potential application value in engineering field.

Great efforts and attempts have been made to explore the dynamics of the RM instability and some comprehensive reviews have been reported, but most focus on single shock \cite{prime2024multiscale,si2024dominant,brasseur2025experimental}. In fact, when a reflected shock wave interacts with an interface, the flow pattern of the fluid and interface morphology may change, leading to more complex physical phenomena. To explore the physical mechanism underlying the RM instability more comprehensively, researchers have investigated the evolution of the RM instability. Ukai \textit{et al.} explored the evolution of the RM instability during the reshock phase under various interface configurations through the compressible Navier--Stokes (NS) equations \cite{ukai2011growth}. Mohaghar \textit{et al.} studied the influence of the initial perturbations on the inclined RM turbulent mixing layer before and after reshock by using the FLASH code \cite{mohaghar2022three}. Based on the Euler equations, Tang \textit{et al.} focused on the single-mode RM instability driven by the convergent shock wave. The results showed that as the Atwood number increases, the perturbation amplitude grows at a faster rate, and the nonlinear phase becomes more prominent \cite{tang2021effect}. Li \textit{et al.} numerically simulated the RM instability of a flat gas interface driven by the perturbed and reflected shock wave, and found that as increase in the Atwood number, both the density gradient and baroclinic vorticity become larger \cite{li2023numerical}. Singh \textit{et al.} investigated the RM instability induced caused by a shock-accelerated square light bubble and observed that enstrophy and dissipation rates increased notably as the shock Mach numbers rose \cite{singh2021contribution}.

Besides the aforementioned studies, fruitful research results have been achieved in theoretical \cite{guo2022large,cong2022experimental,sun2024improved}, experimental \cite{guo2022shock,nagel2022experiments,rasteiro2023effect,guo2024richtmyer} and numerical \cite{wong2019high,groom2019direct,li2019role,latini2020comparison,xiao2020unified,tang2021effects,bin2021new,liang2022phase,wong2022analysis,zhang2022interaction,li2024numerical,schilling2024self} aspects. These results contribute to a comprehensive  understanding of the RM instability. In the RM instability system, there are numerous complex hydrodynamic and thermodynamic non-equilibrium behaviors arise from the multi-wave interaction, nonlinear coupling and multi-scale structures, etc. However, the current researches primarily focus on simulating the dynamic behaviors process of the RM instability system, often ignoring the abundant thermodynamic non-equilibrium (TNE) effects accompanying real-time fluid macroscopic behaviors \cite{shan2018experimental}.

\begin{figure*}[htbp]
	\centering
	\includegraphics[width=0.92\textwidth]{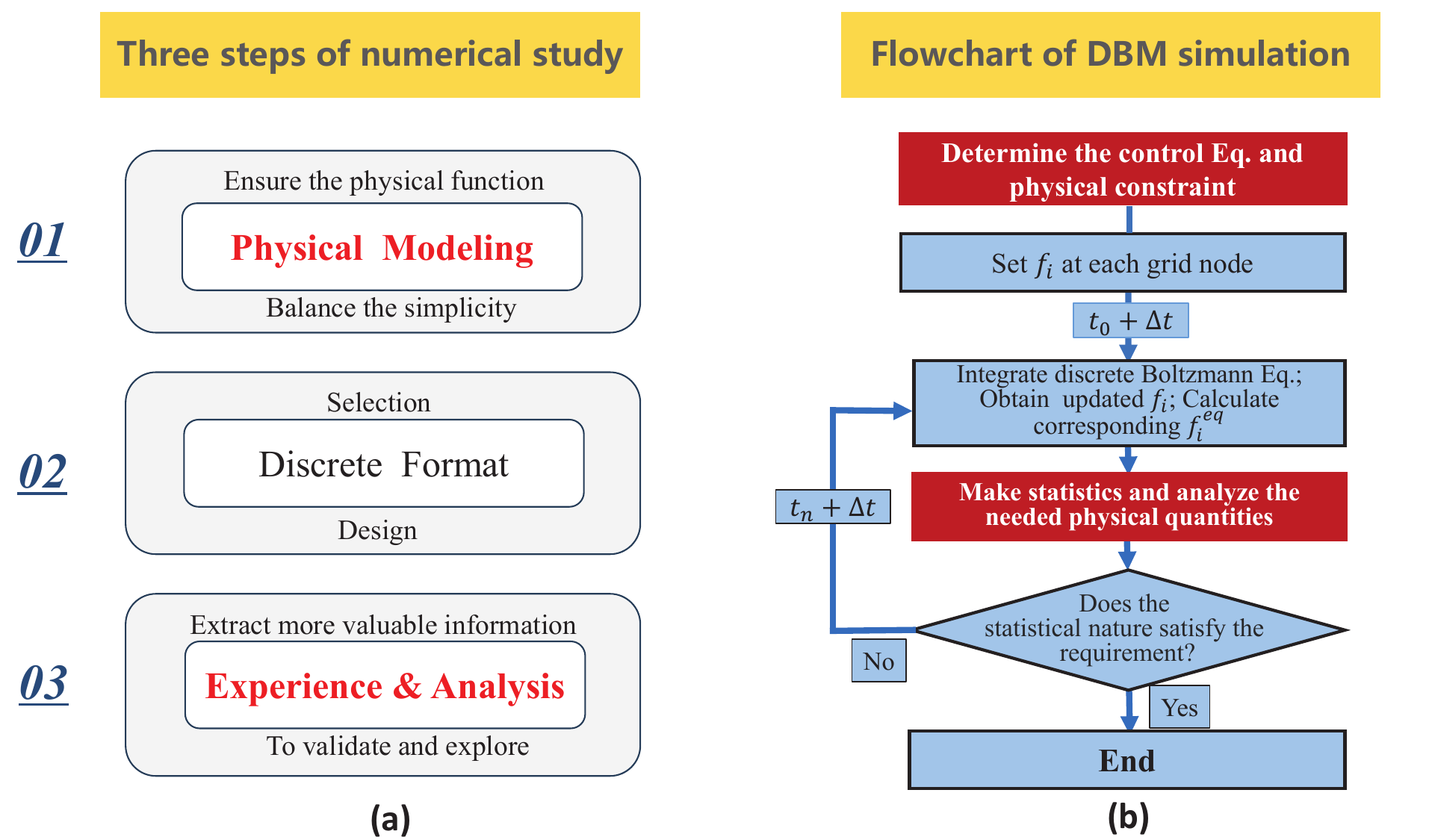}
	\caption{(a) The three main steps of complex flow simulation research.
		(b) Flowchart for DBM simulation.}
	\label{FIG01}
\end{figure*}

To more effectively capture the TNE effects in the RM instability system, the discrete Boltzmann method (DBM) \cite{xu2024advances} is utilized, which is grounded in non-equilibrium statistical physics \cite{Chen2010Nonequilibrium,Xu2022Nonequilibrium,wang2021non}. The DBM can be regarded as a further development or a variation of the lattice Boltzmann method (LBM) \cite{benzi1992lattice,xu2012lattice,wei2018novel,wang2020bounce,wang2020simple,wang2022novel,li2022numerical,chai2023rectangular,li2023equations}. Traditionally, the LBM is mainly used to solve partial differential equations, whereas the DBM serves as a coarse-grained physical model that focuses more on the TNE effects often ignored in macroscopic modeling. In the DBM, through the Chapman--Enskog (CE) multiscale expansion, various orders of the Knudsen number can be used to describe the TNE effects at different levels. Most importantly, the complex TNE effects of the system can be checked, extracted and described based on the non-conservative moments of $(f-f^{eq})$, where $f$ ($f^{eq}$) is the corresponding (equilibrium) distribution function. Therefore, the fluid system can be studied more comprehensively from continuum to near-continuum by utilizing the DBM \cite{Xu2022Nonequilibrium}. Figure \ref{FIG01}(a) shows the three main steps in numerical experimental study, and Fig. \ref{FIG01}(b) illustrates the flowchart outlining the primary steps of DBM simulation. As a modeling and analysis method, DBM is responsible for steps (1) and (3) of three main steps in numerical study shown in Fig. \ref{FIG01}(a) \cite{zhang2023viscous,song2023plasma,xu2024advances}.
%%%%%%%%%%%%%%%%%%%%%%%%%%%%%
%%%%%%%%%%%%%%%%%%%%%%%%%%%%%

At present, the DBM is widely used in fluid systems with significant non-equilibrium behaviors, including multiphase flow \cite{gan2015discrete,zhang2019entropy,wang2023high,sun2022thermodynamic,sun2023droplet},
reactive flow \cite{yan2013lattice,zhang2016kinetic,lin2017multi,lin2018mrt,lin2019discrete,ji2022three,su2023unsteady},
multi-scale compressible flow \cite{gan2018discrete,gan2022discrete}, microscale fluids \cite{zhang2022non,zhang2023lagrangian} , hydrodynamic instability \cite{lin2014polar,lai2016nonequilibrium,lin2016double,de2018discrete,chen2018collaboration,gan2019nonequilibrium,ye2020knudsen, chen2021specific,chen2022discrete,li2022influence,Chen2022,lai2023influences,shan2023nonequilibrium,yang2023influence,song2023plasma,lai2024investigation},
and others systems \cite{liu2023discrete}, providing insights for understanding these fluid systems. In the RM instability system, Lin \textit{et al.} introduced a polar coordinate DBM to simulate the RM instability, where a shock wave travels from a denser (less dense) medium to a less dense (denser) one, and their analysis focused on the non-equilibrium characteristics near both material and mechanical interfaces \cite{lin2014polar}. Subsequently, Lin \textit{et al.} proposed a two-component DBM and simulated the RM instability induced by a detonation wave \cite{lin2016double}. Employing the multiple-relaxation-time DBM, Chen \textit{et al.} explored the non-equilibrium characteristics of a system where RM and Rayleigh--Taylor instabilities coexist, studying the collaboration and competition mechanisms of these two instabilities from a macroscopic perspective \cite{chen2018collaboration}. Started from interpreting the experiment of by Collins and Jacobs (2002, J. Fluid Mech. 464, 113-136) \cite{collins2002plif}, Shan \textit{et al.} investigated the non-equilibrium kinetic effects in the RM instability and reshock process by using the two-fluid DBM, they found the TNE quantities abruptly increase in the shock process and the heavier fluid exhibits higher entropy production \cite{shan2023nonequilibrium}. Based on the characteristics of non-equilibrium behavior, the transmitted shock wave, the compression and dissipation kinetic effects after the shock wave and the material mixing kinetic effects in the later stage are identified. It is found that the viscous stress effect and the heat flux effect caused by the shock wave show very different characteristics: the former is ``transient'', the latter is ``continuous''. Beginning with the analysis of certain laboratory findings, Song \textit{et al.} \cite{song2023entropy} investigated the interaction process between a shock wave and the laminar flow boundary layer on a plate using DBM, mainly focusing on the entropy characteristics and its production mechanism.

Yang \textit{et al.} used the DBM to investigate the RM instability in a two-component system during the reshock process, and found that the higher the density ratio, the higher the TNE quantities and the proportion of the non-equilibrium region for each species, while the heat flux intensity in the heavier medium decreased as the density ratio increased \cite{yang2023influence}. The shock-bubble/droplet interaction is a typical example of RM instability. Starting from interpreting of the experiment of Ding \textit{et al.} in Luo's group \cite{ding2018interaction}, Zhang \textit{et al.} \cite{zhang2023viscous,zhang2023specific} studied the shock wave-bubble/droplet interaction process through a double-distribution function DBM, focusing on the kinetic behavior characteristics that are beyond the ability of traditional fluid theory to describe. Song \textit{et al.} proposed a DBM for plasma kinetics to study the non-equilibrium characteristics in the RM instability, they found that the influence of magnetic field on TNE effect is different before and after interface inversion \cite{song2023plasma}.

In the evolution of the RM instability, the reflection distance plays a crucial role in the development of the interface. Some studies have reported that, with the increase of the reflection distance, the structure of the interface develops more rapidly and transits to the small scale structure under the secondary impact \cite{gowardhan2011numerical,balasubramanian2012experimental,shankar2014numerical}. Wang \textit{et al.} numerically investigated the evolution of the RM instability in a two-dimensional gas cylinder interacted with incident planar shock and reshock, and found that a small reflection distance inhibits the formation of vortex structure, while a larger reflection distance promotes the development of vortex structure \cite{wangxiansheng2012numer}. Reilly \textit{et al.} simulated the incline-interface reshocked RM instability and found that as the reflection distance increases, the more developed the interface, the more energy is deposited by the secondary shock. This is because that the larger density gradient and surface area result in a greater cross product upon reshock \cite{reilly2015effects}. Wang \textit{et al.} explored the effect of reflection distance on the interface instability, and they observed that when the reflection distance increases within a certain range, the post-reshock growth rate of amplitude remains near a constant, and when the reflection distance exceeds a certain value, the growth rate decrease \cite{honghui2022richtmyer}. It is well known that a significant and complex spatial structures emerge during the evolution of the RM instability. Despite these studies, the effect of reflection distance on the RM instability is still unclear. Particularly in terms of how to effectively describe and capture the TNE information of the system, which needs to be further explored. The RM instability in reality is often not as widely studied in the earlier literature: plane shock waves act on the disturbed interface. Earlier studies chose this scenario more often, not because it was true, but because it was simple for initial study. In recent years, the case of curved shock wave interacting with flat material interface and the more general scenarios where a non-uniform disturbance shock wave acting on material interface are attracting increasing attention \cite{zou2020research}.

To better understand the physical mechanism of the RM instability, the effect of reflection distance on the RM instability in the reshock process is investigated through the two-component DBM. The paper is structured as follows: Sec. \ref{SecII} briefly introduces the DBM. Sec. \ref{SecIII} validates the two-component DBM through three benchmarks. Sec. \ref{SecIV} provides a detailed analysis of the influence of the reflection distance on the RM instability and its physical mechanism. Lastly, we give the conclusion in Sec. \ref{SecV}.
\section{Two-component discrete Boltzmann model}\label{SecII}

The two-component discrete Boltzmann model, which is widely used for analyzing the behavior of multi-species systems in fluid dynamics, is based on the discrete Boltzmann equation. The equation for this two-component model takes the following form \cite{lin2016double}:
\begin{equation}\label{e1}
	\frac{{\partial{f_i}^{\sigma}}}{{\partial t}} + \mathbf{v}_{i}\cdot\dfrac{{\partial {f_i}^{\sigma}}}{{\partial {\mathbf{r}}}}=-\dfrac{1}{\tau}({f_i}^{\sigma} - {f_i}^{\sigma eq}),
\end{equation}
where $\sigma$ represents species, ${f_i}^{\sigma}$ (${f_i}^{\sigma eq}$) the discrete (equilibrium) distribution function of each specie, $i$ the index of discrete velocities, $t$ the time, $\mathbf{v}_{i}$ the discrete velocity, ${\mathbf{r}}$ the space coordinate, $\tau=(n^{A}/\theta^{A}+n^{B}/\theta^{B})^{-1}$ the relaxation time in terms of the relaxation parameter $\theta^{\sigma}$ and number density $n^{\sigma}$.

The number density $n^{\sigma}$, mass density $\rho^{\sigma}$, and flow velocity $\textbf{u}^{\sigma}$ for each component are defined as:
\begin{equation}\label{2.3}
	n^{\sigma}=\sum\limits_{i}f_{i}^{\sigma},
\end{equation}
\begin{equation}\label{2.4}
	\rho^{\sigma}=m^{\sigma}n^{\sigma},
\end{equation}
\begin{equation}\label{2.5}
	\mathbf{u}^{\sigma}=\frac{1}{n^{\sigma}}\sum\limits_{i} f_{i}^{\sigma}\mathbf{v}_{i},
\end{equation}
where $m^{\sigma}$ is the molecular mass of component $\sigma$.

The mixing number density $n$, mass density $\rho$, and flow velocity $\mathbf{u}$ can be expressed as:
\begin{equation}\label{2.6}
	n=\sum\limits_{\sigma}n^{\sigma},
\end{equation}
\begin{equation}\label{2.7}
	\rho=\sum\limits_{\sigma}\rho^{\sigma},
\end{equation}
\begin{equation}\label{2.8}
	\mathbf{u}=\frac{1}{\rho}\sum\limits_{\sigma}\rho^{\sigma}\mathbf{u}^{\sigma}.
\end{equation}

The internal energy density of each component and the system's total internal energy density are given by
\begin{equation}\label{2.9}
	E^{\sigma}=\frac{1}{2}m^{\sigma}\sum\limits_{i}{f_i}^{\sigma}\big(|\mathbf{v}_{i} - \mathbf{u}|^{2}+\eta_{i}^{2}\big),
\end{equation}
\begin{equation}\label{2.10}
	E=\sum\limits_{\sigma} E^{\sigma},
\end{equation}
where $\eta_{i}$ accounts for the internal energy from additional degrees of freedom. In this paper, the spatial dimension is defined as $D=2$  and the extra degree of freedom is defined as $I=3$. With these values, the individual temperature $T^{\sigma}$ for each component and the overall mixture temperature $T$ can be calculated as:
\begin{equation}\label{2.11}
	T^{\sigma}=\frac{2 E^{\sigma}}{n^{\sigma}(D+I)},
\end{equation}
\begin{equation}\label{2.12}
	T=\frac {2 E}{n(D+I)}.
\end{equation}
More precisely, the definition of temperature relies on the reference flow velocity, for multi-component mixture, the relationship between DBM and macroscopic hydrodynamic models is not one-to-one, but one-to-several \cite{Xu2022Nonequilibrium,zhang2020two,xu2024advances}.

It should be emphasized that the BGK-like model employed in various kinetic methods, including the DBM, is not the original formulation \cite{bhatnagar1954model} derived from the Boltzmann equation through dynamic simplification. Rather, it represents a modified version of the Boltzmann equation, specifically tailored to certain conditions and integrated with mean field theory. In some aspects, its applicability may extend beyond that of the original Boltzmann equation \cite{xu2024advances,Xu2022Nonequilibrium,gan2022discrete}.

Moreover, it can be shown that the NS equations are recoverable from Eq. \eqref{e1} in the continuum limit (refer to appendix A). This requires that $f^{\sigma eq}_{i}$ satisfies the seven moment relations in Eqs. \eqref{A.1}--\eqref{A.7} (refer to appendix B), which can be represented in matrix form as follows:
\begin{equation}\label{e3}
	\mathbf{C}~ \mathbf{f}^{\sigma eq}=\mathbf{\hat{f}}^{\sigma eq},
\end{equation}
where
\begin{equation}\label{e4}
	\textbf{f}^{\sigma eq}=[f_{1}^{\sigma eq}, f_{2}^{\sigma eq},...,f_{N}^{\sigma eq}]^{\texttt{T}},
\end{equation}
\begin{equation}\label{e5}
	\mathbf{\hat{f}}^{\sigma eq}=[\hat{f}_{1}^{\sigma eq}, \hat{f}_{2}^{\sigma eq},...,\hat{f}_{N}^{\sigma eq}]^{\texttt{T}},
\end{equation}
with a $N \times N$ matrix $\textbf{C}=[\textbf{C}_{1}, \textbf{C}_{2},...,\textbf{C}_{N}]^{\texttt{T}}$, the matrix of coefficients about the discrete velocity $\mathbf{v}_{i}$ and the free parameter $\eta_{i}$. The specific elements of the $\textbf{C}$ are as follows, ${{C}_{1i}}=1$, $C_{2i}=v_{ix}$, $C_{3i}=v_{iy}$, $C_{4i}=v_{i}^{ 2}+\eta _{i}^{2}$, $C_{5i}=v_{ix}^{2}$,
$C_{6i}=v_{ix}v_{iy}$, ${{C}_{7i}}=v_{iy}^{2}$, $C_{8i}=\left( v_{i}^{2}+\eta _{i}^{2} \right)v_{ix}$, $C_{9i}=\left( v_{i}^{2}+\eta _{i}^{2} \right)v_{iy}$,
$C_{10i}=v_{ix}^{3}$, $C_{11i}=v_{ix}^{2}v_{iy}$, $C_{12i}=v_{ix}v_{iy}^{2}$, $C_{13i}=v_{iy}^{3}$,
$C_{14i}=\left( v_{i}^{2}+\eta _{i}^{2} \right)v_{ix}^{2}$, $C_{15i}=\left( v_{i}^{2}+\eta _{i}^{2} \right)v_{ix}v_{iy}$, $C_{16i}=\left( v_{i}^{ 2}+\eta _{i}^{2} \right)v_{iy}^{2}$.
From Eq. \eqref{e3}, the equilibrium distribution function of each component in discrete form can be calculated by
\begin{equation}\label{e6}
	\mathbf{f}^{\sigma eq}= \textbf{C}^{-1} \mathbf{\hat{f}}^{\sigma eq},
\end{equation}
where \(\textbf{C}^{-1}\) represents the inverse of the matrix \(\textbf{C}\), provided that the inverse exists.

An important point to note is that the collision term on the right-hand side of Eq. \eqref{e1} describes the relaxation of the distribution function $f_{i}^{\sigma}$ towards its local equilibrium distribution $f_{i}^{\sigma eq}$. The deviation of the distribution function from its equilibrium counterpart during the relaxation process serves as the source of TNE. The second term on the left-hand side of the equation, often referred to as the convection term, incorporates contributions from both hydrodynamic non-equilibrium (HNE) and TNE effects within the NS hierarchy. To better understand the effects of non-equilibrium, the distribution function $f_{i}^{\sigma}$ is decomposed into its equilibrium and non-equilibrium components as $f_{i}^{\sigma}=f_{i}^{\sigma eq}+f_{i}^{\sigma neq}$. Substituting this decomposition, Eq. \eqref{e1} can be rewritten as:
\begin{equation}\label{TNEandHNE}
	\frac{{\partial{f_i}^{\sigma}}}{{\partial t}} + \mathbf{v}_{i}\cdot\dfrac{{\partial {f_i}^{\sigma eq}}}{{\partial {\mathbf{r}}}}=-\left(\dfrac{1}{\tau}+\mathbf{v}_{i}\cdot\dfrac{\partial }{{\partial {\mathbf{r}}}}\right) {f_i}^{\sigma neq}.
\end{equation}
Indeed, HNE effects arises entirely from the spatial partial derivative of the distribution function on the left-hand side of Eq. \eqref{TNEandHNE}. The CE multiscale analysis indicates that HNE is driven by the gradient of macroscopic quantities. In contrast, TNE is primarily determined by the non-equilibrium component on the right-hand side of the equation, reflecting the system's non-equilibrium behavior when the interactions between microscopic particles fail to fully restore equilibrium. It is important to note that if $f_{i}^{\sigma neq}$ approaches zero, the convection term neglects viscosity and heat conduction. At the Euler level, the discrete Boltzmann equation contains only HNE effects.

\begin{figure}[htbp]
	\centering
	\includegraphics[width=0.30\textwidth]{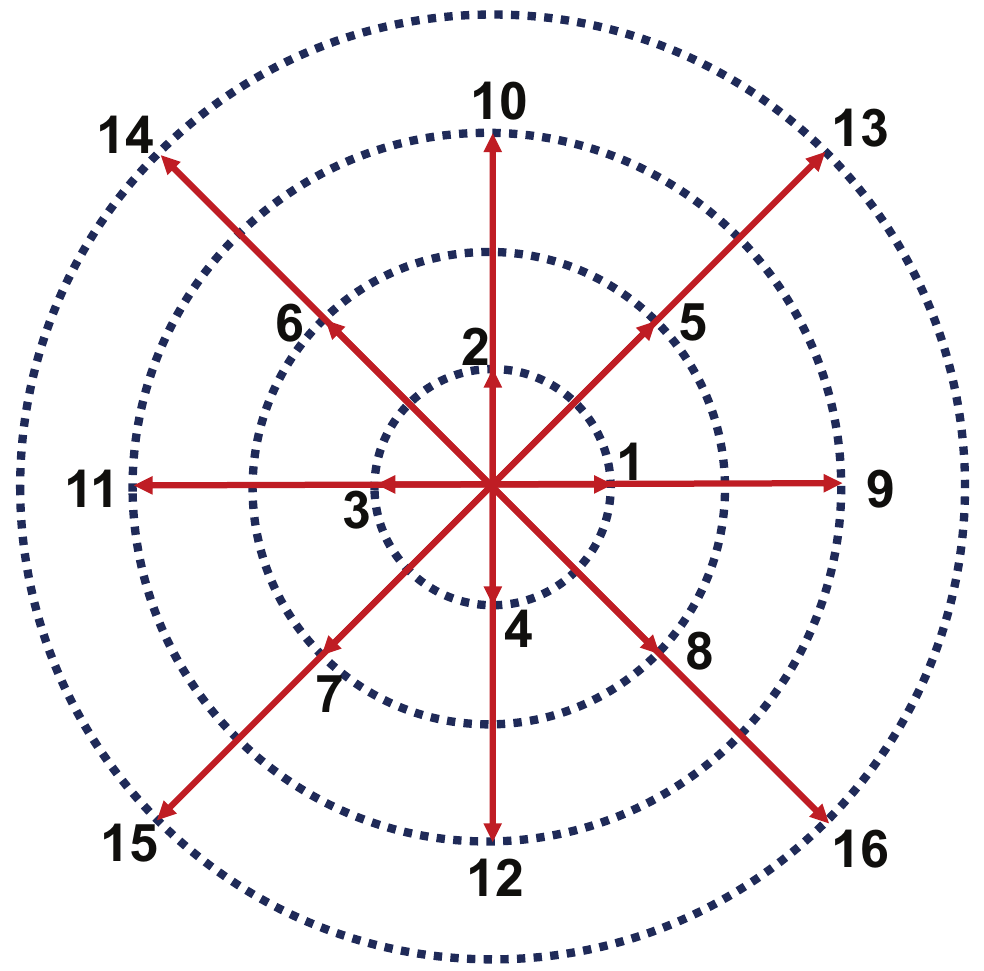}
	\caption{Sketch of D2V16 discrete velocity selecting method.}
	\label{FIG02}
\end{figure}

In this paper, the discrete velocity set D2V16, as shown in Fig. \ref{FIG02}, is used as the discrete velocity space model, and its expression is as follows \cite{lin2019discrete}:
\begin{equation}\label{e7}
	\textbf{v}_{i}=\left\{
	\begin{aligned}
		&\mathbf{cyc}: v_a(\pm1,0), 1 \le {i} \le 4 ,\\
		&v_b(\pm1,\pm1),5 \le {i} \le 8 ,\\
		&\mathbf{cyc}:v_c(\pm1,0),9 \le {i} \le 12 ,\\
		&v_d(\pm1,\pm1),13
		\le {i} \le 16 ,\\
	\end{aligned}
	\right.
\end{equation}
and
\begin{equation}\label{e8}
	\eta_{i}=\left\{
	\begin{aligned}
		&\eta_{a}, 1 \le {i} \le 4 ,\\
		&\eta_{b}, 5 \le {i} \le 8 ,\\
		&\eta_{c}, 9 \le {i} \le 12 ,\\
		&\eta_{d}, 13\le {i} \le 16 ,\\
	\end{aligned}
	\right.
\end{equation}
where $\mathbf{cyc}$ indicates the cyclic permutation, parameters $v_a$, $v_b$, $v_c$, $v_d$ and $\eta_a$, $\eta_b$, $\eta_c$, $\eta_d$ are adjustable. Please note that the discrete formats, including the discrete velocity set, used in the paper are surely sufficient for the current work, but are not necessarily the standard or optimal choice for every scenario. The selection of the optimal discrete velocities depends not only on the discrete time integration and spatial derivative schemes, but also on the specific fluid behavior being studied. This is remains an open question in computational mathematics and falls beyond the scope of this article.

%%%%%%%%%%%%%%%%%%%%%%%%%%%%%
%%%%%%%%%%%%%%%%%%%%%%%%%%%%%

It is noteworthy that $f_{i}^{\sigma eq}$ can be replaced by $f_{i}^{\sigma }$ in Eqs. \eqref{A.1}-\eqref{A.3} according to the conservation of mass, momentum and energy. Therefore, the macroscopic physical quantities (e.g., velocity, density and temperature) of each component can be calculated according to Eqs. \eqref{A.1}-\eqref{A.3}. However, for the Eqs. \eqref{A.4}-\eqref{A.7}, if $f_{i}^{\sigma eq}$ is substituted with $f_{i}^{\sigma}$, the values on both sides of the equation may differ. This deviation of the non-conservation moments of $f_{i}^{\sigma }$ and $f_{i}^{\sigma eq}$ are used to measure the non-equilibrium degree of the fluid system. Based on this, the non-equilibrium quantities are formulated as:

\begin{equation}\label{mn01}
	\bm{\Delta}^{\sigma*}_{m,n}=m^{\sigma}\sum\limits_{i}(f_{i}^{\sigma}-f_{i}^{\sigma eq}) \left(\textbf{v}^{*}_{i}\cdot \textbf{v}^{*}_{i}+\eta^{2}_{i}\right)^{\left(m-n\right)/2}\underbrace{\textbf{v}^{*}_{i}\textbf{v}^{*}_{i}\cdots \textbf{v}^{*}_{i}}_{n},
\end{equation}
where $\bm{\Delta}^{\sigma*}_{m,n}$ represents the contraction of an $m$th-order tensor into an $n$th-order tensor, and  $\mathbf{v}_{i}^{*}=\mathbf{v}_{i}-\mathbf{u}$ is the central velocity.  

To provide a more comprehensive characterization of non-equilibrium effects during the RM instability process, it is advisable to introduce a non-equilibrium strength vector, namely:
\begin{equation}\label{eS}
	\mathbf{S}_{\text{TNE}}=\big\{|\boldsymbol\Delta_{2}^{\sigma *}|,|\boldsymbol{\Delta}_{3,1}^{\sigma\ast }|,%
	|\boldsymbol{\Delta}_{3}^{\sigma\ast }|,|\boldsymbol{\Delta}_{4,2}^{\sigma\ast }|,...,|\boldsymbol{\Delta}%
	_{m,n}^{\sigma\ast }|,{\mid\boldsymbol\Delta^{\sigma *}\mid}\big\},
\end{equation}
which consists of several independent non-equilibrium quantities and their nonlinear combinations. Here, ${\mid\boldsymbol\Delta^{\sigma *}\mid}=\sqrt{\sum_{m,n}|\boldsymbol{\Delta}_{m,n}^{\sigma\ast }|^2}$ is the total TNE quantity. This vector effectively captures the various ways, extents, and processes through which the system deviates from equilibrium, providing a comprehensive, multi-perspective description \cite{sun2023droplet}. 

Furthermore, the average TNE intensity is calculated by first integrating the total TNE quantity and then averaging it across the entire computational domain
\begin{equation}\label{e18}
	\overline{D}^{\sigma *}=\frac{1}{L_xL_y}\int_0^{L_x}\int_0^{L_y} \mid\boldsymbol\Delta^{\sigma *}\mid dxdy,
\end{equation}
where $L_{x}$ and $L_{y}$ denote the length and width of the fluid system, respectively.

To facilitate practical calculations, it is both convenient and advantageous to adopt dimensionless variables. In this study, physical quantities are represented in nondimensional forms  by employing the following reference values, i.e., the molar mass $m_{\theta}$, molar number density $n_{\theta}$,  length $L_{\theta}$, temperature $T_{\theta}$, and universal gas constant $R$. The corresponding nondimensional variables are summarized in Table I. 
%%%%%%%%%%%%%%%%%%%%%%%%%%%%%%%
\begin{table}[ht]\label{TableI}
	\centering
	\caption{Nondimensional variables used in this study.}
	\rule{0pt}{15pt}
	\begin{tabular}{@{}l@{\hspace{3em}}c@{}} 
		\toprule
		\textbf{Physical Quantity and Symbol} & \textbf{Reference Value}  \\ \toprule
		Dsitribution funcitions: $f_{i}^{\sigma}$   & by $n_{\theta}$   \\ \vspace{0.3em}
		Mass density: $\rho^{\sigma}$, $\rho$        &by $m_{\theta}n_{\theta}$       \\  \vspace{0.3em}
		Speed and velocity: $\textbf{v}_{i}$, $\eta_{i}$, $\textbf{u}^{\sigma}$, $\textbf{u}$  &by$\sqrt{RT_{\theta}/m_{\theta}}$       \\ \vspace{0.3em}
		Energy density: $E^{\sigma}$, $E$   & by $n_{\theta}RT_{\theta}$   \\  \vspace{0.3em}
		Temperature: $T^{\sigma}$, $T$        &by $T_{\theta}$       \\  \vspace{0.3em}
		Time: $t$   & by $L_{\theta}/\sqrt{RT_{\theta}/m_{\theta}}$   \\  \vspace{0.3em}
		Distances: $L_{x}$, $L_{y}$, $L_{0}$        &by $L_{\theta}$ \\  
		\bottomrule
	\end{tabular}
\end{table}
%%%%%%%%%%%%%%%%%%%%%%%%%%%%%%%

\section{Validation and verification}\label{SecIII}

In this section, the binary diffusion, sod shock tube and thermal Couette flow are simulated to verify the effectiveness of the DBM model. For the time derivative, the first-order forward Euler scheme is applied, and for the spatial derivatives, the second-order nonoscillatory and nonfree-parameter dissipation difference scheme \cite{zhang1991nnd} is used.

\subsection{Binary diffusion}\label{subsec1}

Initially, the two-component DBM is employed to simulate the binary diffusion phenomenon occurring at the interface between two miscible substances. In the isothermal diffusion process, the initial physical field is as follows:
\begin{equation}\label{Initial}
	\left\{
	\begin{aligned}
		&(n^{A},n^{B})_{L}=(0.95,~0.05), \\
		&(n^{A},n^{B})_{R}=(0.05,~0.95),
	\end{aligned}
	\right.
\end{equation}
where the subscript $L$ denotes $-0.05 \leq L \leq 0.0$ and $R$ denotes $0 < R \leq 0.05$. The calculation region is divided into uniform grids, and the number of grids is selected as $N_x \times N_y =200 \times 1$, the space step $\Delta x=\Delta y=5.0\times 10^{-4}$, the time step $\Delta t=1.0\times 10^{-5}$, the relaxation parameter $\tau^{A}=\tau^{B}=1.0\times 10^{-3}$, the sizes of discrete velocities $(v_{a}, v_{b}, v_{c}, v_{d})=(0.7, 1.5, 1.6, 3.8)$, and parameters $(\eta_{a}, \eta_{b}, \eta_{c}, \eta_{d})=(0.0, 1.0, 0.0, 0.0)$. Furthermore, periodic boundary conditions are applied to the top and bottom boundaries, while a second-order extrapolation scheme is implemented for the left and right boundaries.

According to Fick's law, the analytical solution for the
concentration of each component in the isothermal diffusion process can be expressed as:
\begin{equation}\label{analysic}
	M^{\sigma}=\frac{n_{L}^{\sigma}+n_{R}^{\sigma}}{2}+\frac{\Delta M^{\sigma}}{2} {\rm{erf}} \Big ( \frac{x}{\sqrt{4D_{d}t}}\Big),
\end{equation}
where $\Delta M^{\sigma}$ indicates the initial concentration difference, $ \rm{erf}$ represents the error function and $\sqrt{4D_{d}t}$ represents the diffusion length. In this simulation, the analytical solution of the concentration difference and diffusion coefficient are selected as $\Delta M^{\sigma}=0.9$ and $D_{d}=0.001$ respectively.
%%%%%%%%%%%%%%%%%%%%%%%%%%%%%%%
\begin{figure}[htbp]
	{\centering
		\includegraphics[width=0.39\textwidth]{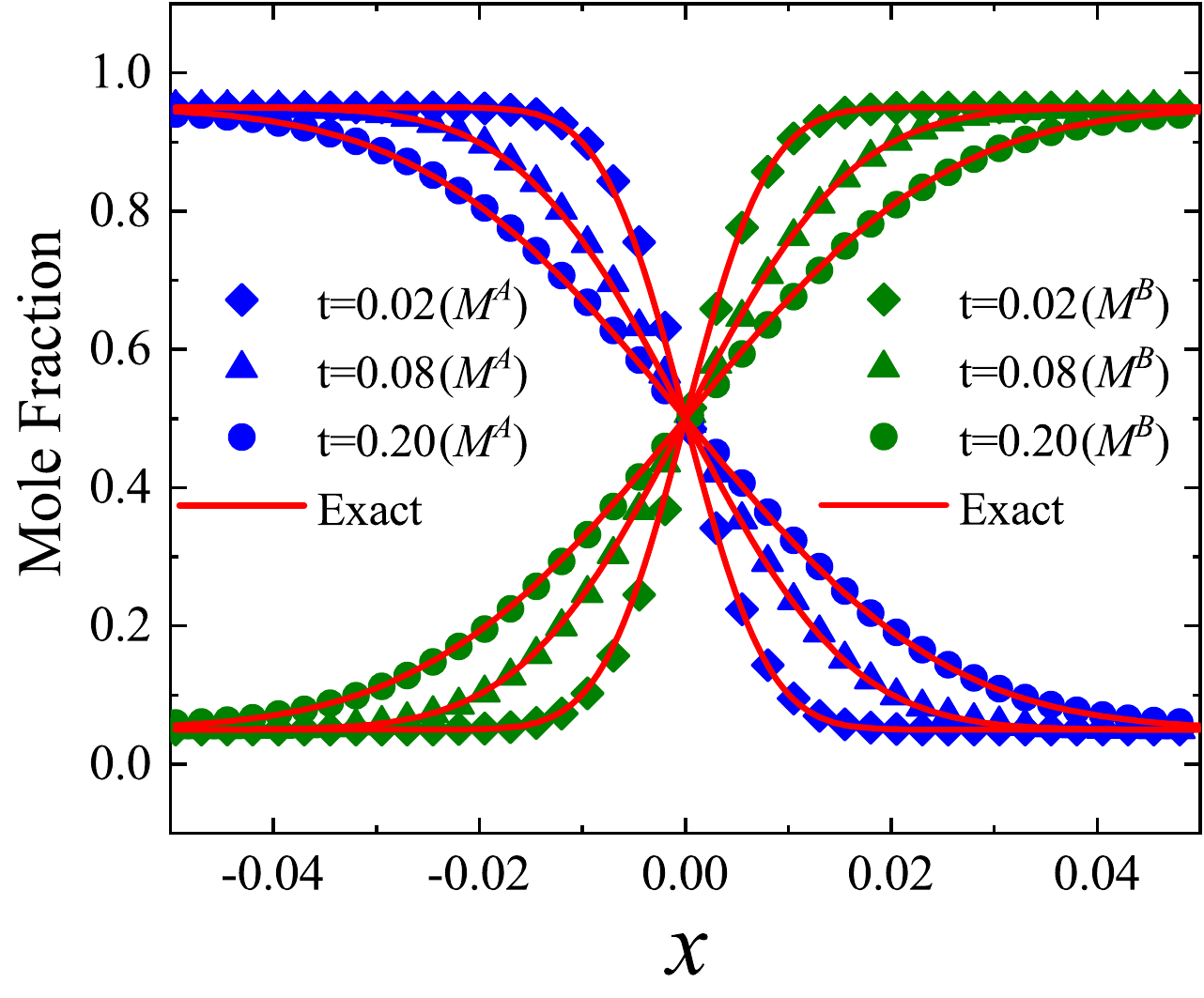}}
	\caption{Binary diffusion process: molar fractions of components $M^{A}$ and $M^{B}$ at three different moments $t = 0.02, 0.08$ and $0.20$, respectively. Symbols represent DBM results, while solid lines depict corresponding analytical solutions.}
	\label{FIG03}
\end{figure}
%%%%%%%%%%%%%%%%%%%%%%%%%%%%%%

Figure \ref{FIG03} displays the distribution of mole fractions for each species, comparing the DBM results with analytical solutions at different time instances $t = 0.02$, $0.08$, and $0.20$, respectively. The symbols represent the DBM results and the solid lines represent the analytical solutions. There is an excellent agreement between the results of simulation and analysis. The results indicate that the DBM can effectively describe the fluid system of two components.

\subsection{Sod shock tube}\label{subsec2}

It is known that the Riemann problem is generally regarded as a classical problem for testing the ability of models in capturing the shock wave. In this subsection, the simulation of the sod shock tube is conducted by using the DBM. The starting configuration of the physical field is as follows
\begin{equation}\label{InitialConfiguration1}
	\left\{
	\begin{aligned}
		&(\rho^{A},T,u_{x},u_{y})\mid_{L}=(1.0, 1.0, 0.0, 0.0),~x \leq 0,\\
		&(\rho^{B},T,u_{x},u_{y})\mid_{R}=(0.125, 0.8, 0.0, 0.0),~x >0,
	\end{aligned}
	\right.
\end{equation}
where the subscripts $L$ and $R$ represent the physical quantities on the left and right sides away from the discontinuous interface, respectively. The computational grid is set to $N_x \times N_y =2000 \times 1$, with the space step of $\Delta x=\Delta y=1.0\times 10^{-3}$ and the time step of $\Delta t=1.0\times 10^{-5}$, the relaxation parameter $\tau^{\sigma}=1.0\times 10^{-5}$. The discrete velocity sizes are chosen as $(v_{a}, v_{b}, v_{c}, v_{d})=(0.6, 1.7, 5.8, 2.8)$ and parameters $(\eta_{a}, \eta_{b}, \eta_{c}, \eta_{d})=(0.0, 2.8, 0.0, 0.0)$.
Furthermore, supersonic inflow boundary conditions are applied in the horizontal direction, while periodic boundary conditions are employed in the vertical direction.

%%%%%%%%%%%%%%%%%%%%%%
\begin{figure}[htbp]
	{\centering
		\includegraphics[width=0.49\textwidth]{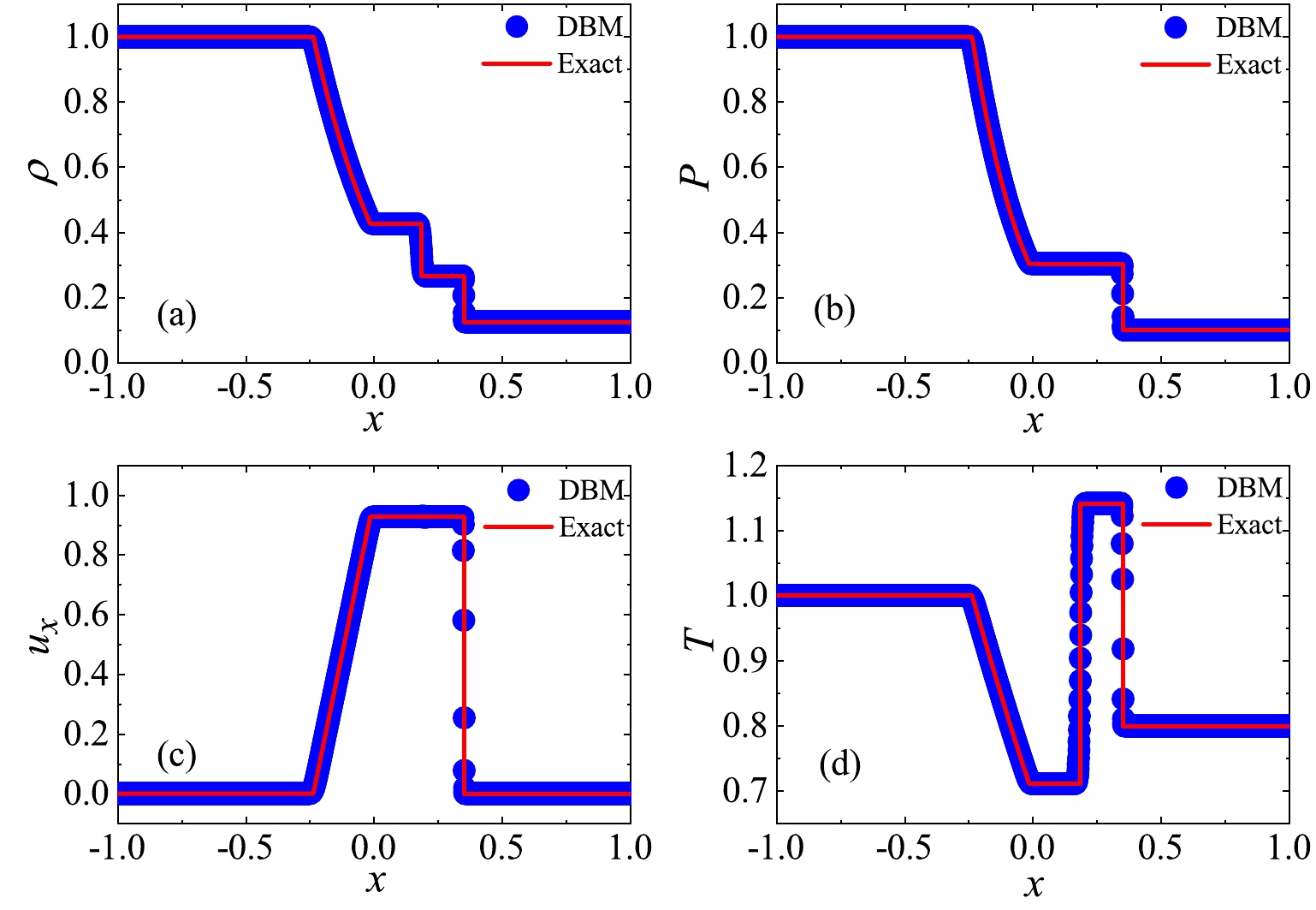}}
	\caption{Profiles of density (a), pressure (b), horizontal velocity (c) and temperature (d) in the Sod shock tube at the moment $t=0.2$. Simulation results are denoted by symbols, while the Riemann solutions are represented by red lines.}
	\label{FIG04}
\end{figure}
%%%%%%%%%%%%%%%%%%%%%%%

The profiles of the density, pressure, horizontal velocity, and temperature at the time $t=0.2$ are displayed in Fig. \ref{FIG04}. The symbols represent the DBM results, while the solid lines correspond to the Riemann analytical solutions. It can be clearly found that a left traveling rarefaction wave, a right traveling shock wave and the material interface between the two components can be captured by the DBM. Furthermore, it is observed that there is a satisfactory consistency between the simulation outcomes and the Riemann analytical solutions, affirming the effective capability of the DBM in capturing the characteristics of the shock wave.

\subsection{Thermal Couette flow}\label{subsec3}

Finally, we perform a simulation of the thermal Couette flow to verify that the DBM can be applied to simulate the compressible fluid with a flexible specific heat ratio. The components $\sigma = A$ and $B$ are located between two parallel plates with a distance of $H=0.1$. At the beginning, the concentrates of the species are $(n^{A},n^{B})=(0.8,0.2)$, molar mass of each species $m^{\sigma}=1.0$, mixing density $\rho=1.0$, mixing temperature $T_{0}=1.0$, and flow velocity $\mathbf{u}^{\sigma}=0.0$. The upper plate moves in the horizontal direction with $u_{0}=0.8$ and the lower plate remains stationary. The non-equilibrium extrapolation scheme is used for the top and bottom, and the periodic boundary conditions are used for the left and right. In addition, the simulation grid is chosen as $N_x \times N_y =1 \times 200$, space step $\Delta x=\Delta y=5.0\times 10^{-4}$, temporal step $\Delta t=1.0\times 10^{-5}$, relaxation parameter $\tau^{\sigma}=1.0\times 10^{-3}$. The parameters for D2V16 are selected as $(v_{a}, v_{b}, v_{c}, v_{d})=(0.8, 1.2, 2.2, 1.1)$ and $(\eta_{a}, \eta_{b}, \eta_{c}, \eta_{d})=(0.0, 1.0, 0.0, 0.0)$.

%%%%%%%%%%%%%%%%%%%%%%
\begin{figure}[htbp]
	{\centering
		\includegraphics[width=0.49\textwidth]{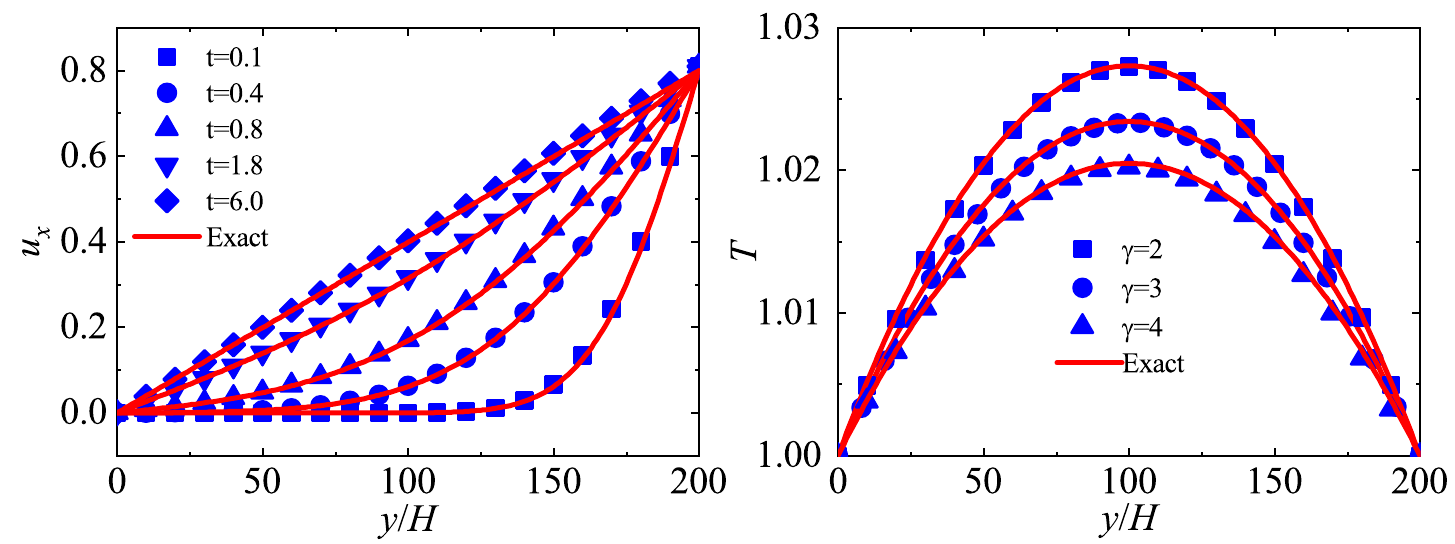}}
	\caption{(a) Horizontal velocity profiles for thermal Couette flow with a specific heat ratio $\gamma$ = 7/5. (b) Temperature profiles for thermal Couette flow with different specific heat ratios: $\gamma$ = 3/2, 7/5, and 4/3, respectively. Symbols denote DBM results, and red solid lines depict analytical solutions.}
	\label{FIG05}
\end{figure}
%%%%%%%%%%%%%%%%%%%%%%%

Figure \ref {FIG05}(a) illustrates the profiles of horizontal velocity $u_{x}$ with specific heat ratio $\gamma = 7/5$ at different moments $t=1.0, 4.0, 8.0, 20$ and $60$, respectively. Here the symbols indicate the results of the DBM and solid lines denote the analytical solutions. The expression of the analytical solution is given by
\begin{equation}\label{couette-u-analysic}
	u=\frac{y}{H}u_{0}+\frac{2}{\pi}u_{0}\sum^{\infty}_{j=1}\Big[\frac{(-1)^{j}}{j}\exp \Big(-j^{2}\pi^{2}\frac{\mu t}{\rho H^{2}}\Big)\sin\Big(\frac{j\pi y}{H}\Big)\Big],
\end{equation}
with $\mu =\tau \rho T$ indicates the dynamic viscosity. Clearly, it can be found that the DBM results match with the analytical solutions.

When the thermal Couette flow reaches the steady state, the theoretical temperature solution is given as follows
\begin{equation}\label{couette-T-analysic}
	T=T_{0}+\frac{u_{0}^{2}}{2c_{p}}\cdot\frac{y}{H}\Big(1-\frac{y}{H}\Big),
\end{equation}
with $c_{p}={\gamma}/({\gamma -1})$. As shown in Fig. \ref{FIG05}(b), the temperature profiles in the vertical direction are considered for three specific heat ratios, i.e., $\gamma =3/2, 7/5$ and $4/3$, respectively. From Fig. \ref{FIG05}(b), it can be observed that the temperature in the vertical direction coincide with the analytical solutions for various specific heat ratios. These results indicate that the DBM has an ability to describe the compressible fluids with adjustable specific heat ratio.

\section{Numerical simulations}\label{SecIV}

%%%%%%%%%%%%%%%%%%%%%%%%%%%%%
\begin{figure*}[htbp]
	\centering
	\includegraphics[width=0.92\textwidth]{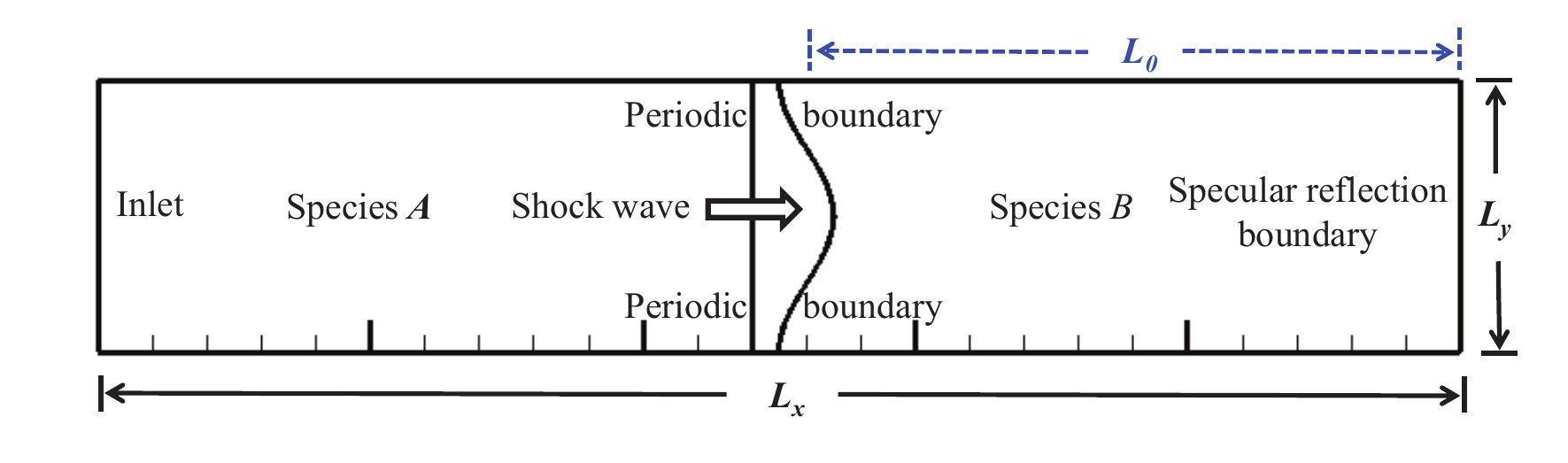}
	\caption{Initial configuration of the RM instability.}
	\label{FIG06}
\end{figure*}
%%%%%%%%%%%%%%%%%%%%%%%%%%%%%

In this section, the RM instability during the reshock process is simulated using the two-component DBM. The initial configuration of the RM instability is illustrated in Fig. \ref{FIG06}. The flow field is divided into three distinct regions. Specifically, the left and middle regions are filled with species $A$, while the right region is entirely composed of species $B$. The physical computational domain is $L_{x}\times L_{y}=0.5 \times 0.1$, and the shock wave propagates through the interface from left to right. A perturbation function is introduced at the interface: $x=0.125+A_{0}\cos (ky)$, where $A_{0}=0.01$ represents the perturbation amplitude and $k=2\pi/L_{y}$ is the wave number. The pressure and temperature on both sides of the material interface are consistent, thus ensuring that the fluid system is in mechanical and thermal equilibrium in its initial state. The concentrations of the two species are selected as $n^{A}=n^{B}=1$, then $m^{A}=1.0$, $m^{B}=5.0$.
In addition, periodic boundary conditions are implemented along the $y$ direction, whereas the left boundary employs inflow conditions and the right boundary uses specular reflection conditions. The initial field is configured as follows:
\begin{equation}\label{InitialConfiguration2}
	\left\{
	\begin{aligned}
		&(\rho,u_{x},u_{y},p)_{L}=(1.3416, 0.3615, 0.0, 1.5133),\\
		&(\rho,u_{x},u_{y},p)_{M}=(1.0, 0.0, 0.0, 1.0),\\
		&(\rho,u_{x},u_{y},p)_{R}=(5.0, 0.0, 0.0, 1.0),
	\end{aligned}
	\right.
\end{equation}
where the subscripts $L$, $M$ and $R$ represent the left, middle and right regions, respectively. The physical quantities across the shock wave satisfy the Rankine-Hugoniot relations \cite{zel2002physics}. Additionally, the relaxation parameters are set as $\theta^{A}=\theta^{B}=2.0\times 10^{-5}$, the time step $\Delta t=5.0\times 10^{-6}$, and the discrete velocity parameter parameters as $(v_{a}, v_{b}, v_{c}, v_{d})=(0.6, 1.6, 2.9, 5.9 )$, free parameter as $(\eta_{a}, \eta_{b}, \eta_{c}, \eta_{d})=(0.0, 2.9, 0.0, 0.0)$. For the sake of effective and accurate simulations, a grid-independent test has been performed first, details can be found in Appendix C. As a result, we choose a grid number of $N_{x}\times N_{y}=2000 \times 400$, which corresponds to the space step of $\Delta x=\Delta y=2.5\times 10^{-4}$ in this simulation. The current work focuses on investigating the effect of the reflection distance on the RM instability during reflected process. This distance is varied by adjusting the position of the shock wave and the material interface. Here, the distance between the shock wave and interface remains a constant, and eight cases with various reflection distances $L_{0}$ are considered, namely, $L_{0}=0.12, 0.16, ... , 0.40$.

\subsection{Flow field visualization}

%%%%%%%%%%%%%%%%%%%%%%%%%%%%%
\begin{figure*}[htbp]
	\centering
	\includegraphics[width=0.79\textwidth]{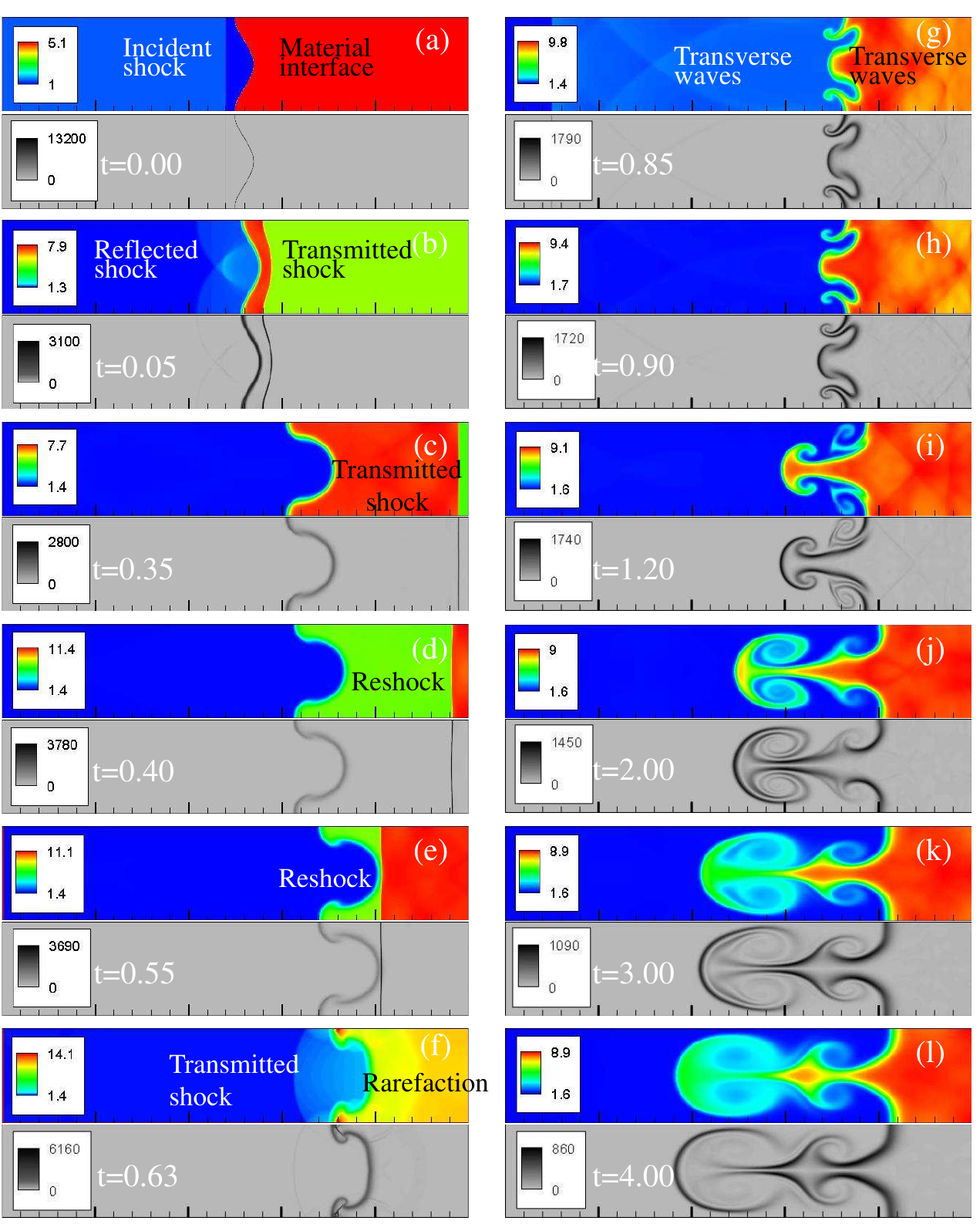}
	\caption{Density contours (top) and schlieren images (bottom) at different moments with a reflection distance of $L_{0}=0.24$.}
	\label{FIG07}
\end{figure*}
%%%%%%%%%%%%%%%%%%%%%%%%%%%%%

To facilitate a clear and intuitive understanding of the evolutionary process of the RM instability, Fig. \ref{FIG07} presents the density contours and schlieren images at various time instances. In the top half of each subplot, the density contours are displayed, while the bottom half shows the schlieren images of density. It can be observed that the RM instability evolves through several distinct stages, including the linear, nonlinear, deformation, and mixing stages. Specifically, in the initial stage, a perturbed interface exists between media $A$ and $B$ (see Fig. \ref{FIG07}(a)). Subsequently, the incident shock wave rapidly traverses the material interface, causing the fluid to enter the linear phase, during which the perturbation amplitude at the interface gradually increases while the interface deformation remains relatively small. Following this, the two species on either side of the material interface begin to diffuse into each other, and ``bubble'' and ``spike'' structures start to form, marking the transition of the fluid from the linear to the nonlinear phase (see Figs. \ref{FIG07}(b)-(d)). At approximately $t = 0.55$, the reshock wave comes into contact with the material interface. Because the reshock here is that the heavier fluid enters the light fluid, producing a left traveling transmitted shock wave and a right traveling reflected rarefaction wave. At this stage, the material interface undergoes a reverse, the peak and valley transform into each other, and is accompanied by the emergence of small-scale structures. Afterwards, under the effect of fluid mixing and velocity shear, the interface widens in the $x$ direction and a noticeable ``spike'' structure appears, the fluid shows a mixing phase (see Figs. \ref{FIG07}(e)-(j)). Finally, when the fluid is adequately mixed, the fine structure begins to disappear and the interface blurs gradually (see Figs. \ref{FIG07}(k)-(l)). The results of these simulations show qualitative agreement with earlier research findings \cite{wanghonghui2022,latini2020comparison}. Moreover, by comparing the density diagrams and schlieren images, it can be observed that a distinct ``mushroom'' structure forms in the heavy fluid as a result of fluid mixing and the baroclinic effect. Simultaneously, due to the continuous penetration of fluids from both sides of the interface, the interface becomes unstable, and a ``plume'' structure emerges. As the interface evolves, these structures gradually develop into ``vortex'' structures and eventually dissipate.

%%%%%%%%%%%%%%%%%%%%%%%%%%%%%
\begin{figure}[htbp]
	\centering
	\includegraphics[width=0.4\textwidth]{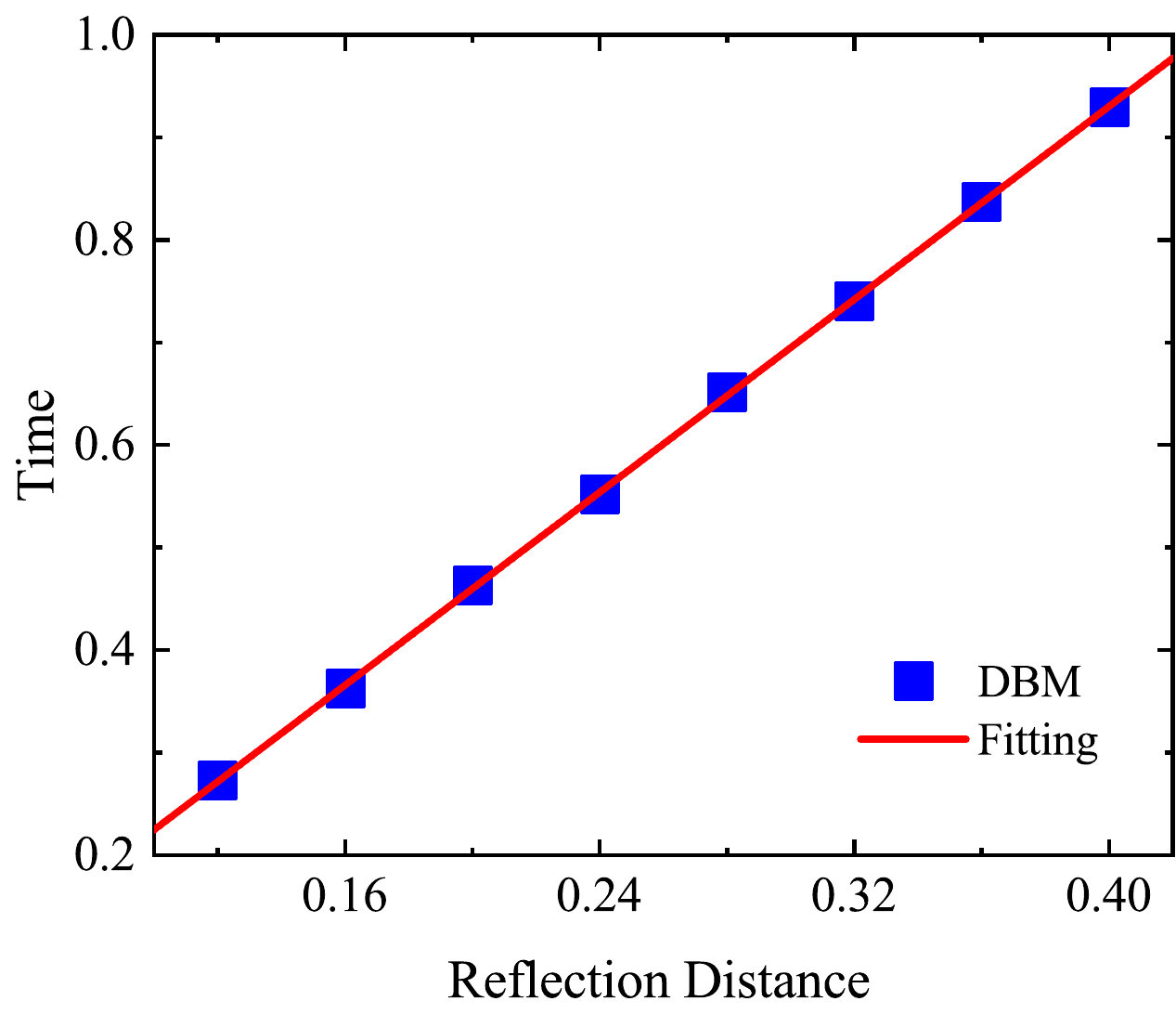}
	\caption{The time when reshock reaches the interface with different reflection distances.}
	\label{FIG08}
\end{figure}
%%%%%%%%%%%%%%%%%%%%%%%%%%%%%

Under varying reflection distances, the evolution of RM instability exhibits significant differences. Figure \ref{FIG08} demonstrates how the interaction time between the reshock wave and the material interface changes with increasing reflection distance. The results show that the secondary interaction time grows linearly with the reflection distance. The corresponding fitting function is given by $L_{0}=2.35t-0.01$, where the slope of 2.35 represents the average velocity of the transmitted shock wave propagating to the right and the reshock wave propagating to the left within component B.

\subsection{Effect of the reflection distance on fluid mixing}

The reflection distance influences the duration of the interaction between the reshock wave and the material interface, thereby affecting the degree of fluid mixing. The following section examines the impact of reflection distance on the RM instability.

\begin{figure*}[htbp]
	\centering
	\includegraphics[width=0.9\textwidth]{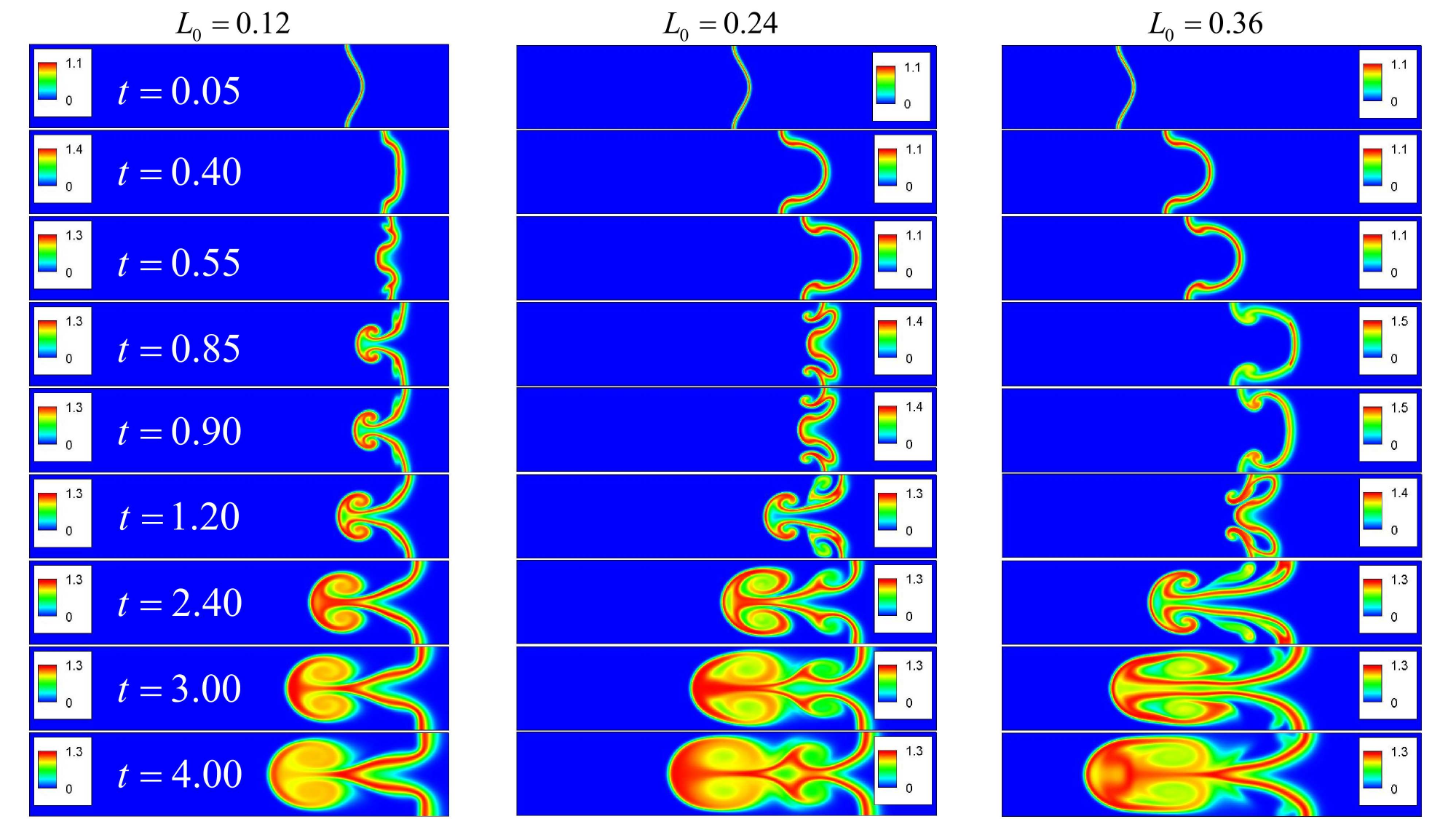}
	\caption{Mixing entropy at different moments with three reflection distances $L_{0} = 0.12, 0.24, 0.36$.}
	\label{FIG09}
\end{figure*}

In thermodynamics, the mixing entropy describes the degree of disorder in a system due to the penetration of two or more species into each other. Mathematically, the expression for the mixing
entropy is
\begin{equation}\label{Entropyofmixing}
	S_{M}=-\sum\limits_{\sigma}n^{\sigma} \ln{X^{\sigma}},
\end{equation}
where $X^{\sigma} = n^{\sigma}/n$ is the mole fraction. Contours of the mixing entropy for three reflection distances are shown in Fig. \ref{FIG09}, where the first to third columns correspond to $L_{0} = 0.12, 0.24$ and 0.36, respectively. It is evident that the mixing area increases with the evolution of the RM instability. This is attributed to the mixing of fluid on both sides of the interface and development of vortex structures that promote the length of the interface, hence extending the mixing region. Moreover, it is evident that the reflection distance substantially influences both the length and shape of the interface. This phenomenon can be attributed to the mixing of fluids on both sides of the interface and the development of vortex structures, which increase the length of the interface and, consequently, extend the mixing region. Furthermore, the reflection distance significantly influences both the length and shape of the interface. Specifically, a larger reflection distance leads to more pronounced changes in the vortex structures and a wider fluid mixing area. This occurs because the reflection distance affects the development of the interface prior to the secondary impact. A more fully developed interface exhibits a larger contact surface when interacting with the reshock wave, thereby intensifying fluid mixing after the secondary impact \cite{reilly2015effects}.

%%%%%%%%%%%%%%%%%%%%%%%%%%%%
%%%%%%%%%%%%%%%%%%%%%%%%%%%%%

\begin{figure*}[htbp]
	\centering
	\includegraphics[width=0.9\textwidth]{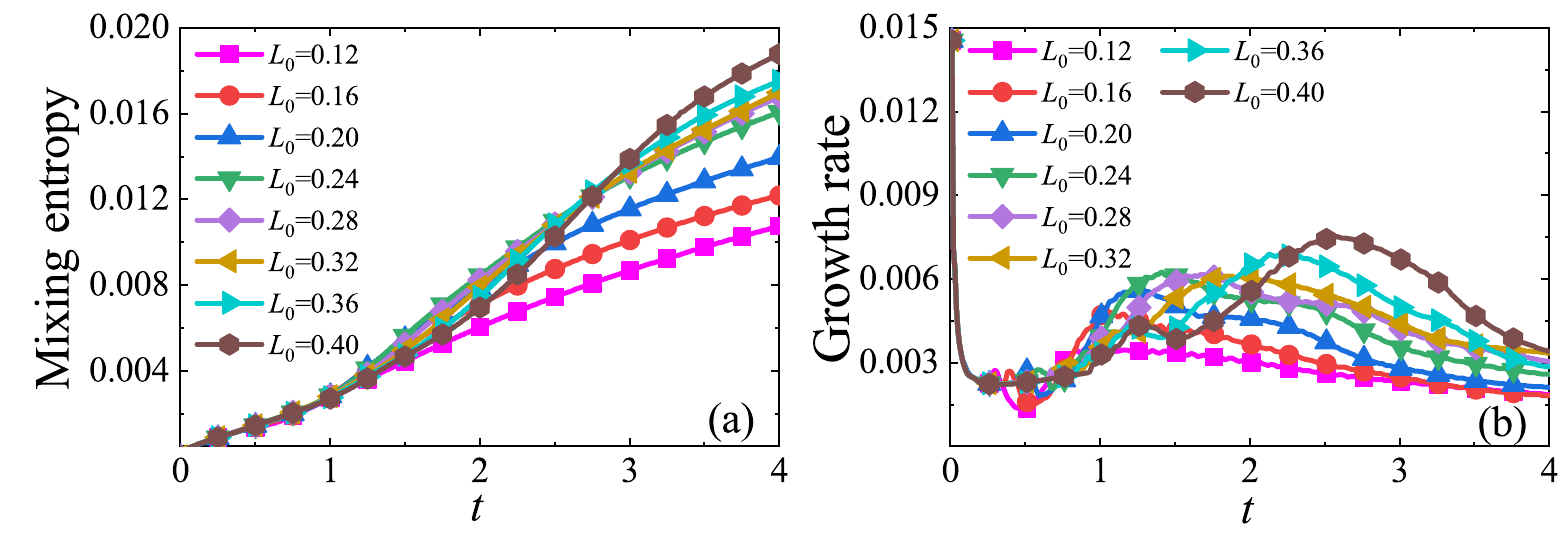}
	\caption{Total mixing entropy (a) and its growth rate (b) in the evolution of the RM instability with various reflection distances.}
	\label{FIG10}
\end{figure*}

To facilitate a quantitative analysis, we present the calculations for the mixing entropy integral $\iint S_{M} dxdy$ along with its growth rat in Fig. \ref{FIG10}, where the integral is extended over the whole region. As seen in Fig. \ref{FIG10}(a), $\iint S_{M} dxdy$ shows a monotonically increasing trend on the whole, and is positively correlated with the reflection distance, which is consistent with the contour of the mixing entropy (see Fig. \ref{FIG09}). To be specific, before $t = 1.0$, there a slight difference in the mixing entropy for different reflection distances, while after $t = 1.0$, a significant difference is demonstrated. In fact, there are the following mixing mechanisms. For those cases, the initial field is the same except that the material interface and the position of the incident shock wave are different. Therefore, the mixing entropy is roughly the same in the early stage until the reshock wave strikes the material interface. And after the secondary impact, as the reflection distance increases, the material interface gets a longer evolution time, resulting in increased density gradients and the contact area of the fluid. This leads to a greater mixing upon reshock.

%%%%%%%%%%%%%%%%%%%%%%%%%%%%%
%%%%%%%%%%%%%%%%%%%%%%%%%%%%%

As illustrated in Fig. \ref{FIG10}(b), the growth rate of $\iint S_M dxdy$ initially decreases, then increases, and eventually decreases. This behavior can be explained by the hybrid mechanisms governing the RM instability. At the beginning, with the decrease of the density difference on both sides of the interface, the mixing rate decreases gradually. Subsequently, the reshock wave accelerates fluid mixing, leading to an increase in the mixing rate. Concurrently, the development of the ``mushroom'' structure expands the fluid's contact area, enhancing diffusion and further accelerating the mixing rate. Finally, as the fluid approaches a fully mixed state, the mixing rate begins to decline.

\subsection{Effect of the reflection distance on HNE characteristic}

%%%%%%%%%%%%%%%%%%%%%%%%%%%%%
\begin{figure*}[htbp]
	\centering
	\includegraphics[width=0.9\textwidth]{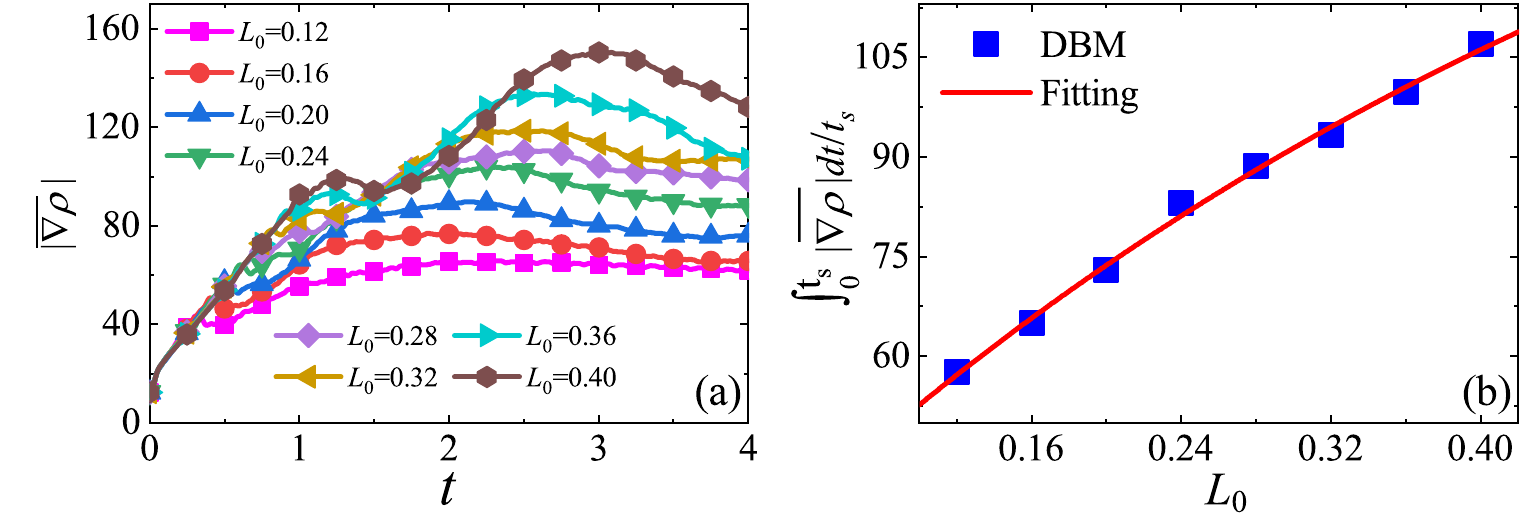}
	\caption{
		(a) Evolution of the average density gradient at varying reflection distances $L_{0}$.
		(b) The relationship between $\int_0^{t_{s}} |\overline{\nabla \rho}|dt/t_{s}$ and $L_{0}$.}
	\label{FIG11}
\end{figure*}
%%%%%%%%%%%%%%%%%%%%%%%%%%%%%

To further investigate the HNE behaviors associated with the RM instability, the average density gradient is analyzed as follows:
\begin{equation}\label{densitygradients}
	|\overline{\nabla \rho}| = \frac{1}{L_xL_y}\int_0^{L_x}\int_0^{L_y} |{\nabla \rho}| dxdy,
\end{equation}
where $|{\nabla \rho}|$ indicates the density gradient and the integral spans the entire computational region $L_{x}\times L_{y}=0.5 \times 0.1$. Figure \ref{FIG11}(a) demonstrates that the evolution of the average density gradient exhibits similar trends across different reflection distances. Specifically, it is characterized by an initial increase, followed by a decrease, a subsequent increase, and a final decline. Moreover, for larger reflection distances, the average density gradient is significantly higher. To elucidate the detailed evolution trend of the average density gradient, we take $L_{0} = 0.24$ as an illustrative example. Before $t=0.55$, the perturbation on the interface gradually amplifies due to the influence of the shock wave, resulting in a continuous widening of the interface and an increase in the integral of the physical gradients. Subsequently, $|\overline{\nabla \rho}|$ rises sharply and then falls rapidly during $t = 0.55\sim0.68$. The increase is attributed to the interaction between the transmitted shock wave and the interface, which enhances the local physical gradients. While the decrease is a result of the interface inversion, the decrease of the interface width and attenuation of the reflected rarefaction wave, leading to the decrease of $|\overline{\nabla \rho}|$. At $t = 0.68\sim2.3$, $|\overline{\nabla \rho}|$ shows an increasing trend. On the one hand, the heavier fluid flows into the lighter fluid, causing an increase in the perturbation amplitude of the interface and accelerating its evolution. On the other hand, the Kelvin-Helmholtz (KH) instability leads to an increase in the width of the interface in the $x$ direction. These two mechanisms contribute to the increase in $|\overline{\nabla \rho}|$. Finally ($t > 2.3$), the fluid structure gradually diminishes due to diffusion and dissipation effects, leading to a decrease in the physical gradients.

Furthermore, to gain a deeper insight into the physical mechanisms of $|\overline{\nabla \rho}|$, the relationship between the reflection distance $L_{0}$ and the time-integrated average density gradient $\int_0^{t_{s}} |\overline{\nabla \rho}|dt/t_{s}$ is illustrated in Fig. \ref{FIG11}(b). The integral is computed over the time range from $t = 0$ to $t_{s}$, where the endpoint is selected as $t_{s}= 4$ for the purposes of this study. The fitting function is given by $\int_0^{t_{s}} |\overline{\nabla \rho}|dt/t_{s}=-623.33\exp(-1.75L_{0})+734.03$. From a physical perspective, as the reflection distance increases, the interface undergoes more extensive development, leading to more pronounced changes in the physical gradients. In contrast, for smaller reflection distances, both the perturbation and the development of the interface are significantly suppressed, resulting in weaker changes in the physical gradients.

\subsection{Effect of the reflection distance on TNE characteristic}

The TNE effects can be explained as the manifestation of molecular thermofluctuations relative to the macroscopic flow velocity $\mathbf{u}$, offering deeper insight into the kinetic influences on the initiation and evolution of RM instability.

%%%%%%%%%%%%%%%%%%%%%%%%%%%%
\begin{figure*}[htbp]
	\centering
	\includegraphics[width=0.9\textwidth]{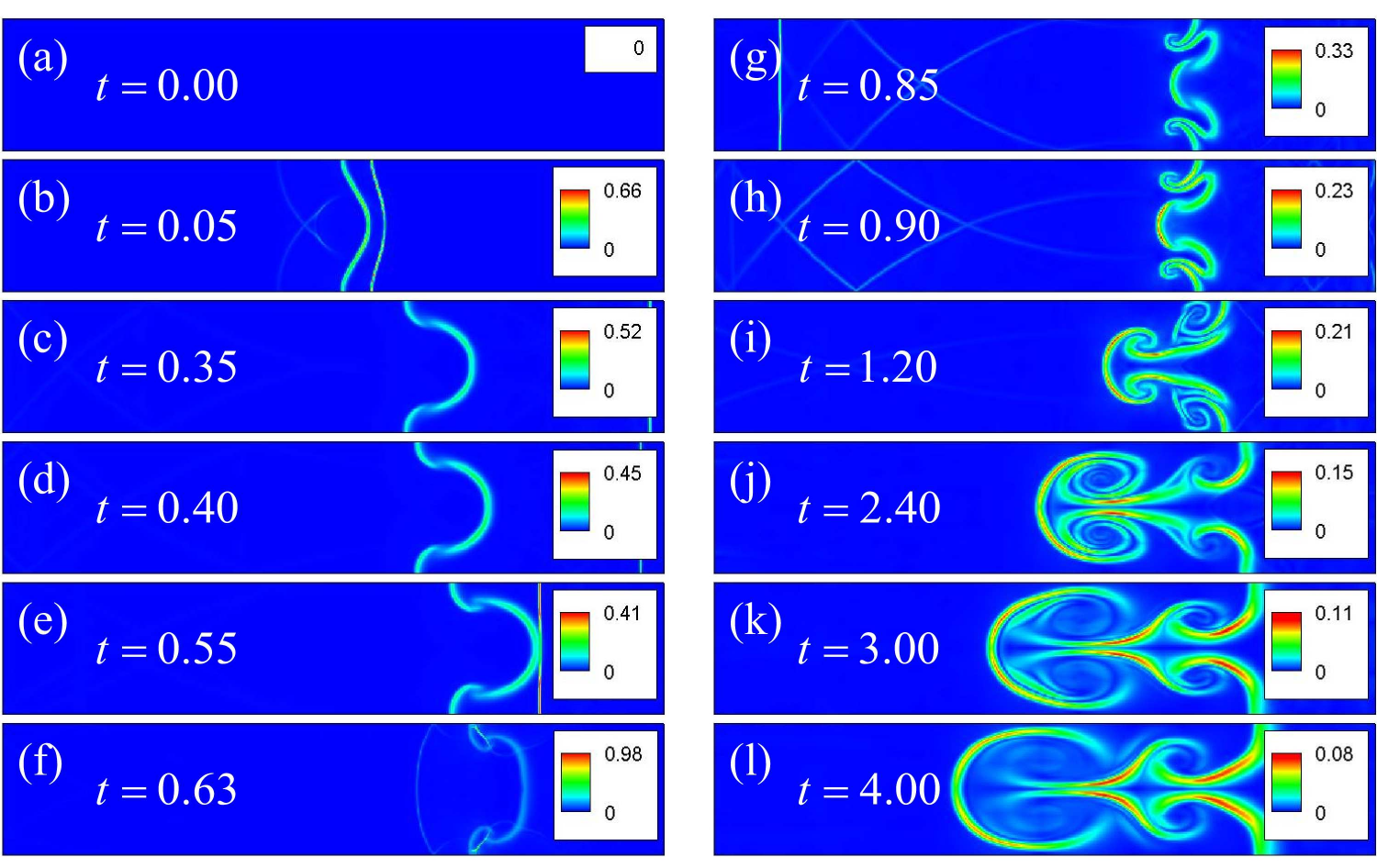}
	\caption{Contours of the total TNE quantity $\mid\boldsymbol\Delta^{\sigma *}\mid$ in the case of $L_{0}=0.24$ at different times.}
	\label{FIG12}
\end{figure*}
%%%%%%%%%%%%%%%%%%%%%%%%%%%%%

For a visual understanding, Fig. \ref{FIG12} depicts the contours of the total TNE quantity $\mid\boldsymbol\Delta^{\sigma *}\mid$ for $L_{0}=0.24$ at different moments. Here the $\mid\boldsymbol\Delta^{*}\mid$  is calculated by the expression as below:
\begin{equation}\label{totalTNEquantity}
	\mid\boldsymbol\Delta^{*}\mid = \mid\boldsymbol\Delta^{A*}\mid+\mid\boldsymbol\Delta^{B*}\mid.
\end{equation}
As shown in Fig. \ref{FIG12}(a), the system initially remains in equilibrium, with $f_i^{\sigma}=f_i^{\sigma eq}$. The non-equilibrium strength increases significantly as the incident shock wave crosses the material interface. Subsequently, as the RM instability progresses, the interface stretches, leading to an expansion of the non-equilibrium region. Ultimately, with sufficient mixing of the two components, the fine structures fade away, and the physical gradients of the system decrease. Furthermore, we can also discover that the non-equilibrium strength is close to zero far from the interface and greater than zero near the interface. This is consistent with the results of the papers \cite{chen2018collaboration,shan2023nonequilibrium}.

%%%%%%%%%%%%%%%%%%%%%%%%%%%%
\begin{figure*}[htbp]
	\centering
	\includegraphics[width=0.9\textwidth]{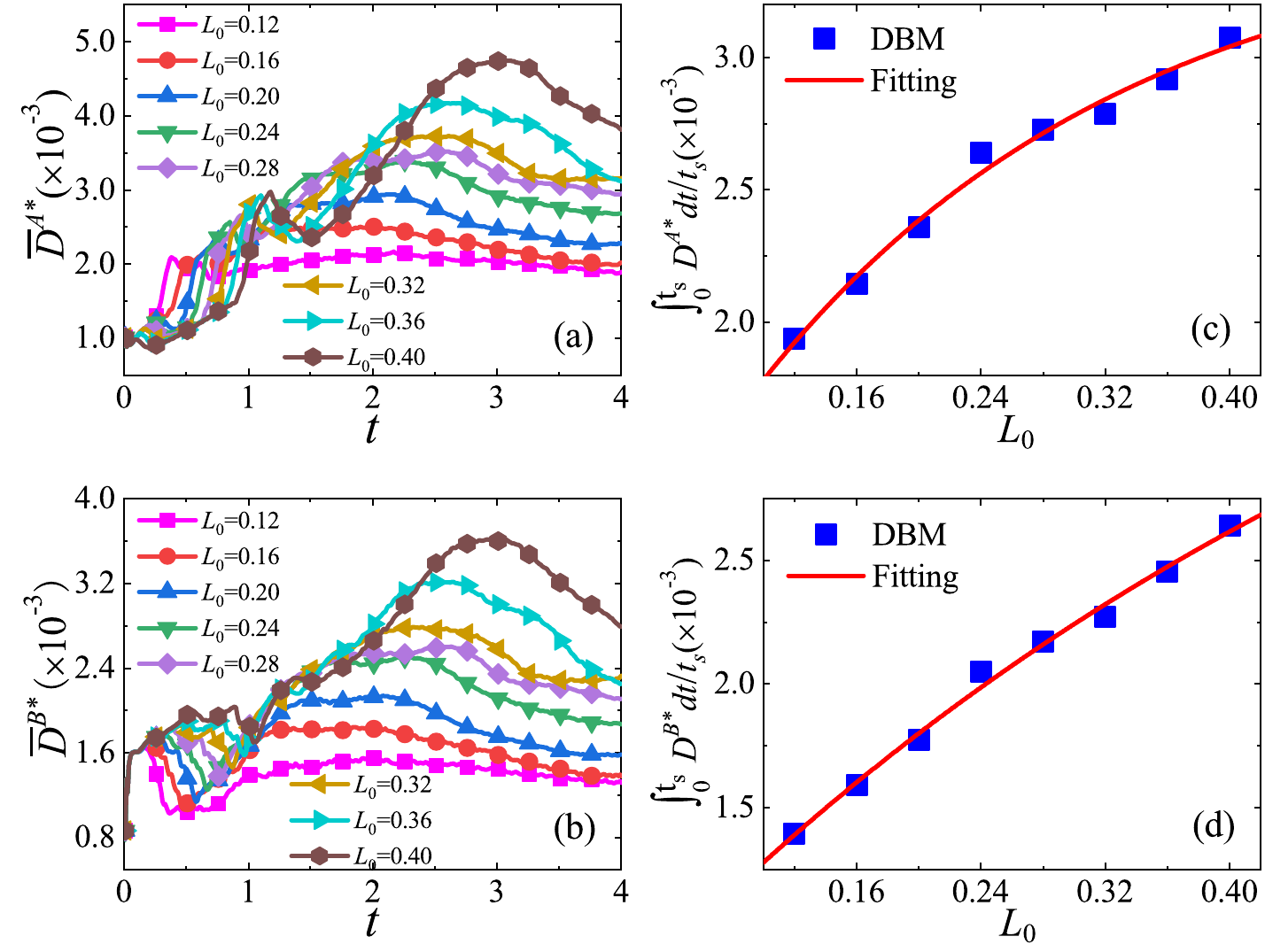}
	\caption{
		(a) Evolution of the average TNE strength of component $A$ with different reflection distances $L_{0}$.
		(b) Evolution of the average TNE strength of component $B$ with different reflection distances $L_{0}$.
		(c) The relationship between $\int_0^{t_{s}} \overline{D}^{A *}dt/t_{s}$ and $L_{0}$.
		(d) The relationship between $\int_0^{t_{s}} \overline{D}^{B *}dt/t_{s}$ and $L_{0}$.}
	\label{FIG13}
\end{figure*}
%%%%%%%%%%%%%%%%%%%%%%%%%%%%%

To conduct a more detailed quantitative analysis of the TNE behaviors of the RM instability, Fig. \ref{FIG13}(a) delineates the evolution of the average TNE strength of the component $A$. On the whole, $\overline{D}^{A *}$ initially rises, then falls, followed by another increase, before ultimately decreasing. Particularly, taking $L_{0}=0.24$ as an example, from $t=0.0$ to 0.25, the stretching of the interface extends the non-equilibrium region, which promotes the rise of $\overline{D}^{A *}$. Subsequently, the $\overline{D}^{A *}$ begins to decrease as the reflected shock wave moves away. At $t=0.3 \sim 0.55$, the transmitted shock wave passes through medium $B$ when the $\overline{D}^{A *}$ change is insignificant. $\overline{D}^{A *}$ demonstrates an upward trend at $t=0.55 \sim 0.9$ when the reshock wave passes through the material interface and becomes a transmitted wave travelling leftwards. Meanwhile, the flow field becomes complex, and the physical gradients of the system increase, hence leads to a rise in $\overline{D}^{A *}$. Later, $\overline{D}^{A *}$ reduces as the departure of the transmitted shock wave and transverse waves exit the computational domain. After $t=1.0$, $\overline{D}^{A *}$ rises again and falls finally. On the one hand, the vortex structure develops and the non-equilibrium region expands due to fluid mixing. On the other hand, the fine fluid structure disappears and the local physical gradients diminish with the adequate mixing of the two species and KH instability. The former (latter) promotes an increase (decrease) in $\overline{D}^{A *}$.

Similarly, Fig. \ref{FIG13}(b) presents the evolution of $\overline{D}^{B *}$ for different reflection distances. It can be seen that, there is a jump increase in non-equilibrium intensity near the initial stage. The system is driven out of equilibrium by the interaction between the incident shock wave and the interface. Later, with the evolution of the RM instability, $\overline{D}^{B *}$ continues to rise. At $t = 0.35 \sim 0.55$, $\overline{D}^{B *}$ exhibits a decreasing trend due to the fact that the transmitted shock wave changes from a curving shape to a plane front, despite the increase in the interface width. In the period of $t=0.55 \sim 0.7$, the reshock wave crosses the material interface, moving out of component $B$, and is accompanied by the weakening of the reflected rarefaction wave, resulting in a tendency to rise again and then fall. Subsequently, the secondary shock promotes the mixing of the fluid, resulting in an increase in the TNE strength. After $t = 2.2$, the evolution of the RM instability enters a later stage of nonlinear development. At this stage, heat conduction and dissipation effects play a dominant role, increasing the mixing degree near the interface and decreasing the physical gradients, consequently causing a reduction in $\overline{D}^{B *}$.

What's more, as shown in Figs. \ref{FIG13}(c) and (d), the relationships between $\int_0^{t_{s}} \overline{D}^{\sigma *}dt/t_{s}$ and reflection distance $L_{0}$ are $\int_0^{t_{s}}\overline{D}^{A*}dt/t_{s}=-0.01\exp(-4.17L_{0})+0.01$ and $\int_0^{t_{s}} \overline{D}^{B *}dt/t_{s}=-0.02 \exp(-1.67L_{0})+0.02$, respectively. It can be found that the TNE intensity is stronger as the reflection distance increases. Physically, this occurs because a greater reflection distance leads to a more rapid evolution of the interface, thereby increasing the degree of mixing after the secondary impact and enhancing the TNE strength.

To provide a more detailed characterization of the TNE strength, we introduce the proportion of the non-equilibrium region $Sr$,  which is defined as the ratio of the non-equilibrium area to the total area of the fluid system \cite{chen2021specific}:
\begin{equation}\label{densitygradients}
	Sr=\dfrac{S^{\sigma}}{S},
\end{equation}
where the $S^{\sigma}$ represents the non-equilibrium region of the component, and $S=L_{x}\times\ L_{y}$ means the computational domain. Specifically, we need to give a threshold value ${\theta}$ in advance. If the value of the TNE strength is greater than the given ${\theta}$, then this point belongs to the non-equilibrium area. In the present work, the threshold value is set at ${\theta}=0.02$.

%%%%%%%%%%%%%%%%%%%%%%%%%%%%
\begin{figure*}[htbp]
	\centering
	\includegraphics[width=0.9\textwidth]{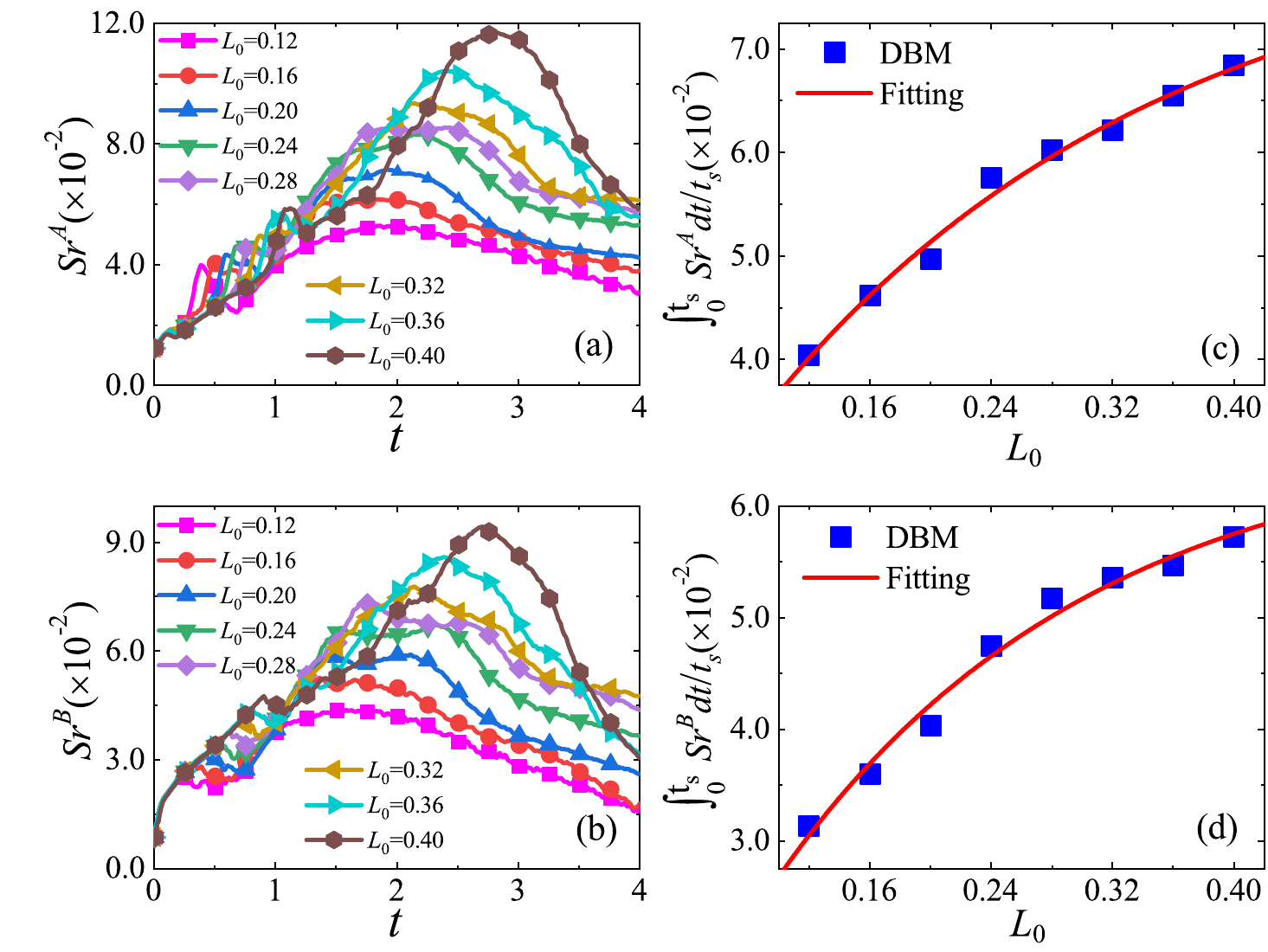}
	\caption{
		(a) Evolution of the non-equilibrium region proportion for component $A$ with various reflection distances $L_{0}$.
		(b) Evolution of the non-equilibrium region proportion for component $B$ with various reflection distances $L_{0}$.
		(c) The relationship between $\int_0^{t_{s}} {Sr}^{A}dt/t_{s}$ and $L_{0}$.
		(d) The relationship between $\int_0^{t_{s}} {Sr}^{B}dt/t_{s}$ and $L_{0}$.}
	\label{FIG14}
\end{figure*}
%%%%%%%%%%%%%%%%%%%%%%%%%%%%%

Figure \ref{FIG14} demonstrates the evolution of $Sr^{\sigma}$ for different reflection distances, exhibiting an initial increase, followed by a decrease, a subsequent increase, and ultimately a decline. Let us take $L_{0}=0.24$ as an example. In Fig. \ref{FIG14}(a), before $t= 0.76$, $Sr^{A}$ rises on the whole due to the elongation of the interface, as well as the effects of the shock wave, reflected rarefaction wave and transverse waves. In the phase of $t=0.76\sim 0.92$, as the transverse waves and transmitted shock wave move out of the computational domain, the local physical gradients of the system decrease. At the same time, the non-equilibrium region decreases due to the interface inversion. These effects lead to a decrease in $Sr^{A}$. During the stage of $t=0.92\sim2.2$, the secondary shock accelerates the evolution of the RM instability, the interface width increases in the $x$ direction, the non-equilibrium region expands, and $Sr^{A}$ shows a fluctuating rise. After $t=2.2$, under the effects of dissipation and heat conduction, the fine structure gradually fade and the system approaches equilibrium, so the $Sr^A$ decreases. As shown in Fig. \ref{FIG14}(b), the evolutionary trend of $Sr^{B}$ is similar to $Sr^{A}$, and the physical mechanism can be revealed through the analysis of $Sr^{A}$. Furthermore, the integration of $Sr^{\sigma}$ over time with various reflection distances is depicted in Figs. \ref{FIG14}(c) and (d). The relationships are given by $\int_0^{t_{s}} {Sr}^{A}dt/t_{s}=-0.27\exp(-3.85L_{0})+0.33$ and $\int_0^{t_{s}} {Sr}^{B}dt/t_{s}=-0.26\exp(-4.76L_{0})+0.27$, respectively. It is evident that as the reflection distance increases, the interface undergoes more extensive development, leading to a more pronounced TNE intensity.

\section{Conclusions}\label{SecV}

In summary, the influence of reflection distance on the RM instability during the reshock process is systematically investigated using a two-component DBM. Various reflection distances are considered, and the behaviors of both HNE and TNE effects are thoroughly analyzed, including the mixing entropy, average density gradient, average TNE intensity, and proportion of the non-equilibrium region. Generally, the interaction time between the reshock wave and the material interface varies with the reflection distance. Moreover, as the reflection distance increases, the fluid system evolves more rapidly, the degree of mixing becomes greater, and the interface structure and TNE behaviors grow more complex. Specifically, it is observed that the reflection distance has a minor effect on the mixing entropy before the secondary impact, but a noticeable difference emerges after the secondary impact. This indicates that the secondary impact significantly enhances fluid mixing. Furthermore, the reflection distance affects the morphology of the vortex structures. A shorter reflection distance suppresses the development of the vortex structure, whereas a longer reflection distance facilitates its growth. Additionally, during the system's evolution, the average density gradient, average TNE intensity, and proportion of the non-equilibrium region exhibit complex variations, characterized by an initial increase, followed by a decrease, a subsequent rise, and a final decline. The time-integrated non-equilibrium quantities and the proportion of the non-equilibrium region increase with larger reflection distances. Physically, the TNE behaviors and macroscopic physical gradients intensify due to the shock wave's impact and interface elongation. Conversely, the non-equilibrium manifestation diminishes due to interface compression, the departure of reflected shock and transverse waves, the attenuation of reflected rarefaction waves, as well as the effects of dissipation/diffusion and heat conduction.

The aforementioned findings deepen our understanding of the RM instability during the reshock process from the perspective of non-equilibrium kinetic effects and provide valuable insights into the underlying mechanisms of RM instability in applied engineering contexts. It is important to note that the present study is limited to cases with relatively low shock wave Mach numbers. As a result, only first-order Knudsen number effects are incorporated into the DBM simulations, ensuring reliable conclusions. For scenarios involving higher Mach numbers, the modeling and simulation using DBM would require the inclusion of higher-order Knudsen number effects. However, due to the significant computational cost associated with such simulations, this aspect will be reserved for future investigation and addressed in a subsequent study.

\begin{acknowledgments}
This work is supported by National Natural Science Foundation of China (under Grant No. U2242214), Guangdong Basic and Applied Basic Research Foundation (under Grant No. 2024A1515010927), Humanities and Social Science Foundation of the Ministry of Education in China (under Grant No. 24YJCZH163), Fujian Provincial Units Special Funds for Education and Research (2022639), Fundamental Research Funds for the Central Universities, Sun Yat-sen University (under Grant No. 24qnpy044), Hebei Outstanding Youth Science Foundation (Grant No. A2023409003), Central Guidance on Local Science and Technology Development Fund of Hebei Province (Grant No. 226Z7601G). This work is partly supported by the Open Research Fund of Key Laboratory of Analytical Mathematics and Applications (Fujian Normal University), Ministry of Education, P. R. China (under Grant No. JAM2405).
\end{acknowledgments}

\appendix
\section{Hydrodynamic equations}\label{A}

Through the CE expansion, it is straightforward to demonstrate the recovery of the compressible NS equations derived from Eq. \eqref{e1} in the continuum limit \cite{lin2016double}:
\begin{equation}
	\label{NSsigma-1}
	\frac{\partial {{\rho }^{\sigma }}}{\partial t}+\frac{\partial }{\partial {{r}_{\alpha }}}\left( {{\rho }^{\sigma }}u_{\alpha }^{\sigma } \right)=0,
\end{equation}%
\begin{eqnarray}
	\label{NSsigma-2}
	&\dfrac{\partial }{\partial t}\left( {{\rho }^{\sigma }}u_{\alpha }^{\sigma } \right)+\dfrac{\partial }{\partial {{r}_{\beta }}}\left( {{\delta }_{\alpha \beta }}{{p}^{\sigma }}+{{\rho }^{\sigma }}u_{\alpha }^{\sigma }u_{\beta }^{\sigma }\right)+\dfrac{\partial }{\partial {{r}_{\beta }}}\left(P_{\alpha \beta }^{\sigma }+U_{\alpha \beta }^{\sigma}\right)\nonumber\\
	&=-\dfrac{\rho^{\sigma}}{\tau^{\sigma}}\left(u_{\alpha}^{\sigma}-u_{\alpha}\right)\text{,}
\end{eqnarray}
\begin{eqnarray}
	\label{NSsigma-3}
	&\dfrac{\partial }{\partial t}\left[{{\rho }^{\sigma }}\left( {{e}^{\sigma }}+\dfrac{1}{2}{{u}^{\sigma 2}} \right)\right]+\dfrac{\partial }{\partial {{r}_{\alpha }}}\left[ {{\rho }^{\sigma }}u_{\alpha }^{\sigma }\left( {{e}^{\sigma }}+\dfrac{1}{2}{{u}^{\sigma 2}} \right)+{{p}^{\sigma }}u_{\alpha }^{\sigma } \right]\nonumber\\
	&-\dfrac{\partial }{\partial {{r}_{\alpha }}}\left[ {{\kappa }^{\sigma }}\dfrac{\partial {T}^{\sigma }}{\partial {{r}_{\alpha }}}-u_{\beta }^{\sigma }P_{\alpha \beta }^{\sigma }+Y_{\alpha }^{\sigma }\right] \nonumber \\
	&=-\dfrac{\rho^{\sigma }}{\tau^{\sigma}}\left(\dfrac{D+I}{2} \dfrac{T^{\sigma}}{m^{\sigma}}+\dfrac{1}{2}u^{\sigma 2}-\dfrac{D+I}{2} \dfrac{T}{m^{\sigma}}-\dfrac{1}{2}u^{\sigma} \right),
\end{eqnarray}
with
\begin{equation}
	P_{\alpha \beta }^{\sigma }=-{{\mu }^{\sigma }}\left( \frac{\partial u_{\alpha }^{\sigma }}{\partial {{r}_{\beta }}}+\frac{\partial u_{\beta }^{\sigma }}{\partial {{r}_{\alpha }}}-\frac{2{{\delta }_{\alpha \beta }}}{D+I}\frac{\partial u_{\chi }^{\sigma }}{\partial {{r}_{\chi }}} \right),
\end{equation}
\begin{equation}
	U_{\alpha \beta }^{\sigma }=-\rho^{\sigma}\left[\dfrac{\delta_{\alpha \beta}}{D+I}\left( u^{\sigma 2}+u^{2}-2u_{\chi}^{\sigma}u_{\chi} \right) \right],
\end{equation}
\begin{equation}
	\begin{aligned}
	Y_{\alpha }^{\sigma}&=\dfrac{\rho^{\sigma}u_{\alpha}^{\sigma}}{D+I}\left( u_{\beta}^{\sigma}-u_{\beta} \right)^{2}+\rho^{\sigma}\left(u_{\alpha}^{\sigma}-u_{\alpha} \right)\\
	&\times\left( -\dfrac{D+I+2}{2}\dfrac{T^{\sigma}-T}{m^{\sigma}}-\dfrac{1}{2}u^{\sigma 2}+\dfrac{1}{2}u^{2}\right),
	\end{aligned}
\end{equation}
where $p^{\sigma }=n^{\sigma}T^{\sigma }$, $e^{\sigma }=(D+I)T^{\sigma }/(2m^{\sigma })$, $\mu ^{\sigma }=p^{\sigma }\tau ^{\sigma }$, $\kappa^{\sigma}=\gamma\mu ^{\sigma }$ and $\gamma=(D+I+2)/(D+I)$ denote the pressure, internal energy per unit mass, dynamic viscosity coefficient, heat conductivity of species $\sigma$ and specific heat ratio, respectively. The $T$ denotes the mixing temperature, $D$ denotes the spatial dimension, $I$ indicates the extra degrees of freedom, $\delta_{\alpha\beta}$ represents the Kronecker function and the subscripts $\alpha, \beta, \chi $ stand for $x$ or $y$. It is crucial to emphasize that the capability to derive the aforementioned hydrodynamic equations represents only one facet of the DBM's physical functionalities.

The physical functionality of the DBM is manifested through its extended hydrodynamic equations. These equations encompass not only the standard three hydrodynamic equations but also incorporate several closely associated non-conservative moment evolution equations. The inclusion of these additional equations, especially the evolution equations for pertinent non-conserved moments, becomes increasingly critical as the system's discreteness or degree of non-equilibrium becomes more pronounced.

\section{Kinetic moment relations}\label{B}

The discrete equilibrium distribution function $f_{i}^{\sigma eq}$ satisfies the following seven moment relations:
\begin{equation}\label{A.1}
	\sum\limits_{i}f_{i}^{\sigma eq}=\iint f^{\sigma eq} d\mathbf{v}d\eta,
\end{equation}
\begin{equation}\label{A.2}
	\sum\limits_{i}f_{i}^{\sigma eq}\mathbf{v}_{i}=\iint f^{\sigma eq}\mathbf{v} d\mathbf{v}d\eta,
\end{equation}
\begin{equation}\label{A.3}
	\sum\limits_{i}f_{i}^{\sigma eq}(\mathbf{v}_{i}\cdot \mathbf{v}_{i}+\eta_{i}^{2})=\iint f^{\sigma eq}(\mathbf{v}\cdot \mathbf{v}+\eta^{2})d\mathbf{v}d\eta,
\end{equation}
\begin{equation}\label{A.4}
	\sum\limits_{i}f_{i}^{\sigma eq}\mathbf{v}_{i}\mathbf{v}_{i}=\iint f^{\sigma eq}\mathbf{v}\mathbf{v}d\mathbf{v}d\eta,
\end{equation}
\begin{equation}\label{A.5}
	\sum\limits_{i}f_{i}^{\sigma eq}(\mathbf{v}_{i}\cdot \mathbf{v}_{i}+\eta_{i}^{2})\mathbf{v}_{i}=\iint f^{\sigma eq}(\mathbf{v}\cdot \mathbf{v}+\eta^{2})\mathbf{v}d\mathbf{v}d\eta,
\end{equation}
\begin{equation}\label{A.6}
	\sum\limits_{i}f_{i}^{\sigma eq}\mathbf{v}_{i}\mathbf{v}_{i}\mathbf{v}_{i}
	=\iint f^{\sigma eq}\mathbf{v}\mathbf{v}\mathbf{v}d\mathbf{v}d\eta,
\end{equation}
\begin{equation}\label{A.7}
	\sum\limits_{i}f_{i}^{\sigma eq}(\mathbf{v}_{i}\cdot \mathbf{v}_{i}+\eta_{i}^{2})\mathbf{v}_{i}\mathbf{v}_{i}=\iint f^{\sigma eq}(\mathbf{v}\cdot \mathbf{v}+\eta^{2})\mathbf{v}\mathbf{v}d\mathbf{v}d\eta,
\end{equation}
where the local equilibrium distribution function $f^{\sigma eq}$ is
\begin{equation}\label{e2}
	f^{\sigma eq}=n^{\sigma}\Big(\frac{m^\sigma}{2\pi T}\Big)^{D/2}\Big(\dfrac{m^\sigma}{2\pi IT}\Big)^{1/2}  \exp \Big(-\frac{m^\sigma\mid\mathbf{v}-\mathbf{u}\mid^2}{2T}-\frac{m^\sigma \eta ^{2}}{2IT}\Big).
\end{equation}

\section{Grid independence}\label{C}

%%%%%%%%%%%%%%%%%%%%%%%%%%%%%
\begin{figure}[htbp]
	\centering
	\includegraphics[width=0.45\textwidth]{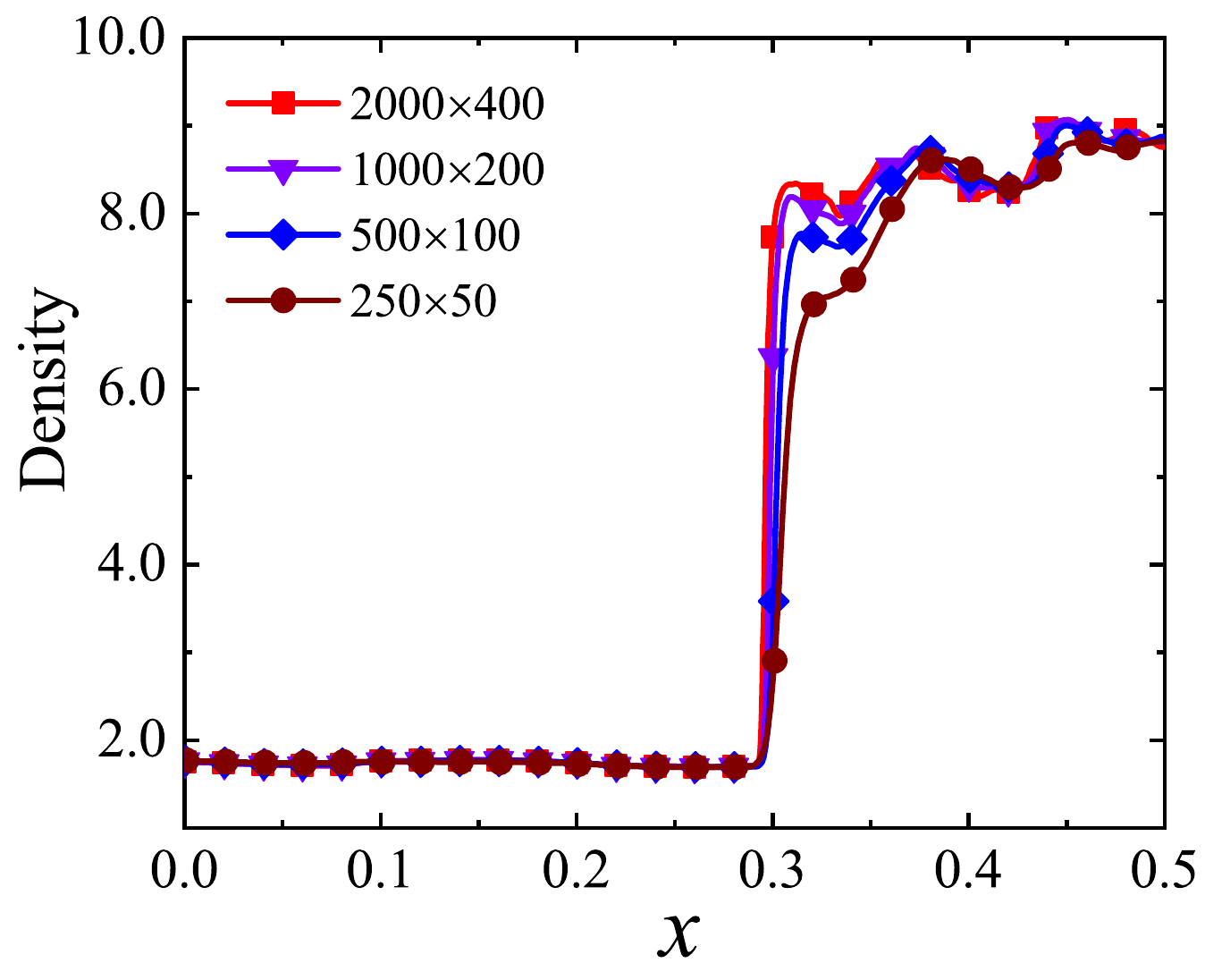}
	\caption{Verification of grid convergence in the RM instability process: In the case of $L_{0}=0.24$, the density profile $\rho$ along the center line $L_{y}/2$ is depicted at time instant $t=1.2$ for four distinct mesh grids.}
	\label{FIG15}
\end{figure}
%%%%%%%%%%%%%%%%%%%%%%%%%%%%%

For the consideration of numerical accuracy, a grid independence test is performed. Four kinds of grid sizes are chosen, $N_x\times N_y=250\times50$, $500\times100$, $1000\times200$ and $2000\times400$, corresponding space steps are $\Delta x=\Delta y=2.0\times 10^{-3}$, $1.0\times 10^{-3}$, $5.0\times 10^{-4}$ and $2.5\times 10^{-4}$, respectively. As shown in Fig. \ref{FIG15}, it can be observed that as the spatial step size decreases, the simulation results converge closely to one another. This demonstrates that the numerical error diminishes progressively as the number of grids increases. Taking account of the numerical resolution and accuracy, a grid size of $N_x\times N_y=2000\times400$ with a space step of $\Delta x=\Delta y=2.5\times 10^{-4}$ is selected to simulate the RM instability.

%This heightened complexity is reflected in the diversity of physical phenomena and the increased nonlinearity of constitutive models and governing equations.

%As a result, the degree of nonlinearity in the system is substantially amplified. %, emphasizing the necessity for improved models.

\section*{Data Availability}
The data that support the findings of this study are available from the corresponding author upon reasonable request.

%\appendix
\section*{References}
\bibliography{DBM2RMI}% Produces the bibliography via BibTeX.

\end{document}